\documentclass[11pt]{article}
\usepackage{amssymb,graphicx,epsfig} 
\evensidemargin=\oddsidemargin 
\def\baselinestretch{1.2}

\def\toR{\raisebox{-.6ex}{\rlap{$\longrightarrow$}} \raisebox{.6ex}{\scriptsize ~$\ \not{\!\!\mathcal{R}}$}~~} 
\def\L{\mathcal{L}}
\def\B{\mathcal{B}}
\def\S{\mathcal{S}}
\def\R{\mathcal{R}}
\def\lprod{\lambda^{\prime *}_{223} \lambda^\prime_{323}}
\def\lamp{\lambda^\prime}
\def\twtwth{\lambda^\prime_{223}}
\def\thtwth{\lambda^\prime_{323}} 
\def\lsim{\raisebox{-.4ex}{\rlap{$\sim$}} \raisebox{.4ex}{$<$}} 
\def\gsim{\raisebox{-.4ex}{\rlap{$\sim$}} \raisebox{.4ex}{$>$}} 
\begin{document}
\begin{titlepage}
\begin{flushright}
SINP/TNP/2011/01  \hfill TIFR/TH/10-37
\end{flushright}

\begin{center}
{\LARGE{\bf
{Can Flavor Physics Hint at Distinctive Signals \\ [2mm]
for R-parity Violation at the LHC?}}} \\[5mm]
\bigskip
{\sf Biplob Bhattacherjee} $^{a,1}$, 
{\sf Gautam Bhattacharyya} $^{b,2}$, 
and
{\sf Sreerup Raychaudhuri} $^{a,3}$ 
\\ [4mm]
\bigskip

\small
{\noindent $^{a)}$ 
Department of Theoretical Physics, 
Tata Institute of Fundamental Research,  \\
\hspace*{0.1in} 1 Homi Bhabha Road, Mumbai 400 005, India. }

\medskip

\noindent $^{b)}$ 
Saha Institute of Nuclear Physics, 
1/AF Bidhan Nagar, Kolkata 700 064, India. 
\end{center}
\normalsize

\vskip 30pt
\begin{center}
{\large\bf ABSTRACT}
\end{center}

\begin{quotation} 
\noindent Observation of some low-energy processes in the flavor physics 
regime may require the existence of supersymmetry with two relatively large 
R-parity-violating couplings of the LQD-type, together with reasonably 
light superparticles. At the LHC, such interactions would be expected to 
give rise to clear signals with convenient leptonic triggers, including 
some multileptons of the same sign. We undertake a detailed 
investigation of these signals taking care to correlate with low-energy 
requirements and taking proper account of the Standard Model backgrounds 
as well as the R-parity-conserving sector of the supersymmetric model. 
We find clear indications that R-parity violation as envisaged in this 
scenario can be detected at the LHC --- even, perhaps, in the early 
runs.
\end{quotation}

\bigskip 

\begin{center} 
PACS numbers: {\tt 11.30.14v, 14.20.Sv, 12.60.Jv} 
\end{center}

\vskip 60pt
\begin{center}
\today
\end{center}

\vfill

\hrule
\vskip -10pt
\footnotesize \noindent
$^1$biplob@theory.tifr.res.in \qquad\qquad
$^2$gautam.bhattacharyya@saha.ac.in \hfill
$^3$sreerup@theory.tifr.res.in
\normalsize
\end{titlepage}
%%%%%%%%%%%%%%%%%%%%%%%%%%%%%%%%%%%%%%%%%%%%%%%%%%%%%%%%%%%%%%%%%%%%%%%%%%%%%%
\newpage
\setcounter{page}{1}

\section{Introduction: Unifying Approaches}

The Standard Model (SM), which represents the culmination of eighty 
years of investigation into the nature and properties of elementary 
particles, can be described as both a triumph and a tragedy. The triumph 
lies in our having been able to put together a stable, working model 
which explains a whole world of empirical data to an accuracy which 
sometimes reaches the amazing value of one in a hundred thousand. The 
tragedy is that the selfsame model is really a portmanteau containing 
disparate, not-so-compatible theoretical ideas and a bunch of purely 
phenomenological parameters, and hence, lacks credibility as a final 
theory. Normally, such an obviously-incomplete theory would be expected 
to fall apart under the impact of systematic empirical study, but the SM 
seems to have passed four decades of stringent experimental testing with 
flying colors \cite{Altarelli}. Much debate and discussion \cite{Quigg} 
during this period has made it quite clear that if we are to bridge some 
of these gaps in our understanding, we must invoke new physics beyond 
the SM, even though all efforts to find empirical evidence for such new 
physics have so far proved futile. It is, therefore, vital to the 
progress of the subject that we continue our search for new physics in 
all possible ways. Moreover, the lesson to be learnt from the failure of 
previous searches is that the task at hand is by no means an easy one 
and hence, we should bring all possible resources to bear upon the 
problem.

In this all-encompassing pursuit of new physics beyond the SM, the bulk 
of terrestrial experimentation is being channelled, at the present 
juncture of time, along two more-or-less parallel approaches. One is the 
set of low-energy experiments probing flavor physics in meson production 
and decay modes, such as the BELLE, BaBar and Fermilab Tevatron 
experiments, where one hopes that as more and more statistics is 
gathered, some unusual flavor-changing processes might be observed at 
rates larger than predicted in the SM. To this category also belong the 
different neutrino experiments probing flavor violation in the leptonic 
sector. The other approach is that of running accelerators at the high 
energy frontier, specifically at CERN's Large Hadron Collider (LHC), 
where the general hope is that new particles may begin to make their 
appearance as the centre-of-mass energy and the luminosity of this 
machine are pushed to hitherto-unattained levels. The former approach is 
sensitive to new physics through virtual effects, i.e. they work because 
of quantum mechanics. Such effects would provide only indirect clues, of 
course, but they have the advantage of not being restricted by the 
machine energy. The latter approach can confirm the existence of new 
physics by direct creation and detection of heavy, unstable particles, 
i.e. this approach works because of special relativity. Direct discovery 
is, of course, the best kind of existence proof, but in this case it is 
severely constrained by the kinematic reach of the LHC machine. We thus 
have two experimental paradigms --- one looking for virtual states and one 
for real states. However, just as quantum mechanics and special 
relativity are not mutually incompatible, neither are the above 
approaches mutually exclusive, and they do not need to be pursued in 
isolation. What seems to be called for is a dexterous amalgamation of the 
two different techniques, one providing hints of the nature of couplings 
and masses of the new physics and the other following those leads in 
order to pin it down completely. More specifically, if there is a model 
of new physics which contributes to low energy processes at levels which 
would soon become testable, it is imperative to check out the possible 
signals of such a model at the LHC; conversely, if a model seems 
accessible to detection at the LHC, one would immediately ask what its 
low-energy consequences are. This article builds on such an idea and 
makes a naive -- but nonetheless robust -- attempt at unifying the two 
approaches.

To be more specific, in this work we consider a model of 
R-parity-violating (RPV) supersymmetry~\cite{rpar} and study its dual 
impact, viz., on the one hand, its influence in flavor physics through 
virtual effects, and on the other, its prediction of specific 
multilepton and jet final states at the LHC leading to possible 
identification of superparticle resonances. As is well known, 
R-parity is a discrete 
symmetry defined by $\R = (-)^{\,3\B+\L+2\S}$, where $\B$, $\L$, and 
$\S$ are, respectively, the baryon number, lepton number and spin of a 
particle: thus we have $\R = +1$ for all SM particles, while $\R = -1$ 
for their superpartners. We begin by noting that conservation of 
R-parity, though a popular ingredient in models of supersymmetry, is not 
demanded by any deep underlying principle. It was
originally introduced by Farrar and Fayet \cite{rpar} to prevent 
fast proton decay through operators which violate lepton and baryon 
number. However, it was later pointed out by Weinberg and others \cite{rpar} 
that it is phenomenologically sufficient to conserve baryon number but not 
lepton number, or vice versa, in order to ensure stability of the proton. 
Most supersymmetric grand unified theories prefer a 'baryon parity' which 
calls for baryon number conservation, because this is 
anomaly-free \cite{Ross}, and hence, most of the literature on RPV tends 
to violate lepton number through either $LL\bar{E}$ or $LQ\bar{D}$ operators. 
Baryon number conservation is, therefore, taken implicitly in our work. 
We then go on to argue, following Ref.~\cite{Bhattacharyya:2009hb}, that the 
simultaneous presence of some specific RPV operators in the superpotential can 
provide correlated enhancements over the SM prediction 
in several flavor-changing neutral 
current (FCNC) channels and thereby predict flavor-changing leptonic 
decays at experimentally-verifiable levels which lie far above the SM
expectations. Fixing the parameters of 
this model to the values demanded by this requirement, we then go on to study 
the impact of such interactions in superparticle production and decays 
at the LHC. It turns out that with these assumptions distinctive and rather spectacular signals are predicted for a wide range of parameters in 
the R-parity-conserving (RPC) sector of the model. The techniques used 
in our analysis, though applied in the context of specific RPV couplings 
as above, are fairly general and 
can be easily adapted to search for any RPV signal due to operators of 
the same type. There is a pleasing synergy, therefore, between flavor physics 
studies such as those of Ref.~\cite{Bhattacharyya:2009hb} and our LHC study, 
viz., discovery of some of the FCNC 
effects predicted at low energies would imply spectacular signals at the 
LHC and discovery of some such signals at the LHC would immediately tell 
us that some spectacular FCNC effects must occur at low energies.

Our work differs from previous collider studies for RPV \cite{RPV_LHC} 
in that we do not allow the masses and couplings relevant to the RPV 
sector to vary freely and independently (as has been done by previous 
workers in the field), but constrain them by the requirement that they 
should predict a certain set of 
low-energy FCNC effects at or close to their present 
upper bounds. In a sense it is this {\it assumption} which directly 
leads to correlation of large new physics signals in low-energy 
processes with 
spectacular LHC signals, and vice versa. It is also worth pointing out that 
some of the signals for RPV at the LHC do not depend very crucially on this 
correlation of couplings -- thus, these would be observed even if we do not 
invoke any low-energy connection.
 
The plan of the article is as follows. In the next section, we summarize 
the arguments set out in detail in Ref.~\cite{Bhattacharyya:2009hb}, 
explaining the requirement for specific terms in the superpotential. We 
then discuss the different possibilities for non-SM LHC signals which 
are thrown up by the presence of these operators in Section~3. Section~4 
is devoted to the study of resonances in these non-SM signals, while 
Section~5 discusses the case when resonances would not be apparent. 
Section~6 discusses same-sign leptons and the discovery limits using 
these signals. In the concluding Section~7 we include a critical summary 
of our results.

%%%%%%%%%%%%%%%%%%%%%%%%%%%%%%%%%%%%%%%%%%%%%%%%%%%%%%%%%%%%%%%%%%%%%%%%%%%%%%
\section{R-Parity Violation: the Flavor Connection}

The fact that large mixing of neutrino flavors ({\it maximal} mixing 
between $\nu_\mu$ and $\nu_\tau$) has been firmly established \cite{pdg} 
has set afoot speculations that lepton flavor-violating (LFV) 
transitions could also be observed in the charged lepton sector. 
However, if this were to proceed in exact analogy with the quark sector, 
then the unitarity of the mixing matrix and the smallness of the 
neutrino masses would jointly ensure that SM predictions for LFV decays 
of charged leptons $\ell_i ~(= e, \mu, \tau)$ would lie far below any 
detectable level\footnote{This is analogous to the 
Glashow-Iliopoulos-Maiani (GIM) cancellation in the quark sector, except 
that the residual effects are much smaller, being suppressed by the tiny 
neutrino masses.}. It follows that actual observation of processes like, 
e.g.,
$$
\ell \to 3 \; \ell' \ ,
\qquad\qquad 
\ell \to \ell' + \gamma \ ,
\qquad\qquad
\ell \to \ell' \, + P \ ,
$$ 
where $P$ is a pseudoscalar meson and $\ell$ and $\ell'$ are leptons of 
different flavour, would immediately and unambiguously establish the 
existence of new physics beyond the SM. These are expected to be probed 
with ever-increasing sensitivity in different ongoing and upcoming 
experiments \cite{Bona:2007qt}. Concerning such new physics, two 
comments are in order, viz.
\vspace*{-0.2in}
\begin{itemize} 

\item The new interaction(s) should be capable of enhancing the LFV 
decay rates, but at the same time keeping the light neutrino 
masses\footnote{Generation of Majorana masses, of course, requires 
lepton number violation by two units.} under control, and

\item If the above decays are observed near their present upper limits, 
then new physics at the tree level must be mediated by particles with 
weak scale couplings and having masses not more than a few hundred GeV.

\end{itemize}
\vspace*{-0.2in}
Clearly, particles of this nature would be copiously produced at the 
LHC. Moreover, differential rates of LFV decays could be directly linked 
to the relative abundance of one lepton flavor over another in observed 
final states at the LHC and therefore, provide prior motivation for 
choosing preferential lepton triggers in the detection process.

In this paper, we concentrate on the $\L$-violating $\lamp$-type 
superpotential, given in terms of chiral superfields, by
\begin{equation}
\widehat{W}= \sum_{i,j,k=1}^3 \  \ \lambda'_{ijk} \, \left(
\widehat{\nu}_{Li} \; \widehat{d}_{Lj} - \widehat{\ell}_{Li} \; \widehat{u}_{Lj} \right)
\widehat{d}_{Rk}^{~c}
\end{equation}
where $i,j,k$ are generation indices. These lead to LFV interactions 
whose strength is determined by the 27 coupling constants 
$\lambda'_{ijk}$. Phenomenological constraints exist on each of these 
$\lambda'_{ijk}$ couplings and also on many of their products, taken 
pairwise \cite{reviews,Kao:2009fg}. However, if we consider all 27 of 
them together the phenomenological situation would become chaotic and it 
would be well-nigh impossible to make definite predictions. It is usual, 
therefore, to work with a small number -- often just one -- of the 
$\lambda'_{ijk}$ couplings, which are assumed to be large, while the 
others are zero or vanishingly small. This kind of scenario is inspired 
by the structure of Yukawa couplings in the SM, which are large only in 
the third generation, while the others are negligibly small by 
comparison. In this work, we select only two of the $\lambda'_{ijk}$ 
couplings, namely $\twtwth$ and $\thtwth$ to be large, i.e., close to 
their present upper limits.

The question immediately arises: why choose just this pair of 
third-generation couplings $\twtwth$ and $\thtwth$ to be large, when 
there are many possible choices?  Before justifying our choice, however, 
let us review the existing constraints on these two couplings.  Limits 
from low-energy processes will always depend on the mass $\widetilde{m}$ 
of the $\widetilde{c}_L$ scalar, or the $\widetilde{b}_R$ scalar. 
Keeping in mind the present direct search limit on squarks from their 
non-observation from the Tevatron experiment \cite{pdg}, we assume a 
benchmark sparticle mass $\widetilde{m} = 300$~GeV for this part of our 
analysis. Then, the best upper limit on $\twtwth$ comes 
from~\cite{reviews,Kao:2009fg,Bhattacharyya:1995pq}
$$
R_{D^0}\equiv \frac{{\rm BR}\;(D^0 \to K^- \mu^+ \nu_\mu)}
                   {{\rm BR}\;(D^0 \to K^- e^+ \nu_e)} \; ,
$$ 
and its value is
\begin{equation}
\twtwth \ \lsim \ 0.3 \left( \frac{\widetilde{m}}{300~{\rm GeV}} \right)
\end{equation}
at 90\% C.L.. On the other hand, the best upper limit on $\thtwth$ 
arises from~\cite{Kao:2009fg}
$$
R_{D_s}(\tau\mu) \equiv \frac{{\rm BR} (D_s^+ \to \tau^+ \nu_\tau)}
                             {{\rm BR} (D_s^+ \to \mu^+ \nu_\mu)} \; ,
$$ 
and the 90\% C.L.value is 
\begin{equation}
\thtwth \ \lsim \ 0.9 \left( \frac{\widetilde{m}}{300~{\rm GeV}} \right)
\end{equation}
These upper bounds of 0.3 and 0.9 respectively (for $\widetilde{m} = 
300$~GeV) are rather high and comparable to the values of the 
electroweak couplings. However, as the mass $\widetilde{m} = 300$~GeV is 
significantly higher than the masses of the electroweak gauge bosons, 
the RPV effects would be weaker than SM effects of the same order. In 
fact, these new physics effects can only dominate processes where the 
same-order SM effects are suppressed, e.g. by a GIM-type mechanism.

We now come to the special role played by $\twtwth$ and $\thtwth$. It 
turns out \cite{Bhattacharyya:2009hb} that the joint action of these two 
couplings can trigger a variety of potentially-observable phenomena, 
such as those listed below.
\vspace*{-0.2in}
\begin{itemize} 

\item \underline{\sl Generation of neutrino masses}: 
Quite generally, neutrino mass terms are generated at the one-loop level 
via $\lamp$ couplings by combinations $\lambda'_{ijk} \lambda'_{i'kj}$, 
which yield $[m_\nu]_{ii'}$. Typically, for a sfermion mass of 300 GeV, 
these products of coupling constants will have an upper limit of 
$10^{-4}$ to $10^{-6}$ \cite{gb}, where the exact limit depends on the 
mass of the fermions of the $j$ and $k$ generations. However, if we take 
$\twtwth$ and $\thtwth$ as the only large ($\sim 0.5$) couplings 
(together with a small $\lambda'_{113} \,\lsim \,0.1$), neutrino masses 
are not generated at one loop, but an acceptable pattern of elements in 
the neutrino mass matrix emerges at the two-loop level 
\cite{Dey:2008ht}. The key point here is that it is essentially the 
two-loop suppression factor that allows $\twtwth$ and $\thtwth$ to be as 
large as their upper limits obtained from other phenomenological 
considerations, provided all the other $\lambda'$ couplings are small.

\item \underline{\sl Higher rates for LFV decays}: 
The simultaneous presence of couplings $\twtwth$ and $\thtwth$ induces 
LFV decays of the $\tau$ lepton, viz.
$$
\tau \to 3\mu \;, \qquad 
\tau \to \mu\gamma \;, \qquad
\tau \to \mu \eta  \;, \qquad
\tau \to \mu \eta' \; ,
$$
at numerically significant rates. The scalars that mediate these 
processes are $\widetilde{b}_R$ and $\widetilde{c}_L$. As before, we 
denote the mass of a generic scalar by $\widetilde{m}$. We also define a 
`product coupling' $\lamp~\equiv~\sqrt{\lprod}$. The numerical 
contributions to different branching ratios and their dependence on 
$\lamp$ and $\widetilde{m}$ are given by \cite{Bhattacharyya:2009hb}:
\begin{eqnarray} 
\label{brtau}
{\rm BR}\, (\tau \to 3\mu) &\simeq& 3.9 \times 10^{-7} 
\left[1.0 - 0.6 \left(\frac{\lamp}{0.7}\right)^2 +
0.1 \left(\frac{\lamp}{0.7}\right)^4\right] 
\left(\frac{\lamp}{0.7}\right)^4 
\left(\frac{300~{\rm GeV}}{\widetilde{m}}\right)^4 \, ; \nonumber \\
{\rm BR}\, (\tau \to \mu\gamma) &\simeq& 1.0 \times 10^{-6} 
\left(\frac{\lamp}{0.7}\right)^4 
\left(\frac{300~{\rm GeV}}{\widetilde{m}}\right)^4 \, ; \nonumber \\
{\rm BR}\, (\tau \to \mu\eta) &\simeq& 3.4 \times 10^{-7} 
\left(\frac{\lamp}{0.7}\right)^8 
\left(\frac{300~{\rm GeV}}{\widetilde{m}}\right)^4 \, ; \nonumber \\
{\rm BR}\, (\tau \to \mu\eta^\prime) &\simeq& 3.3 \times 10^{-7} 
\left(\frac{\lamp}{0.7}\right)^8 
\left(\frac{300~{\rm GeV}}{\widetilde{m}}\right)^4 \ .
\end{eqnarray} 
These may be compared with the present experimental upper limits of the 
above branching ratios, as listed by the Particle Data Group \cite{pdg}, 
which are
\begin{eqnarray}
{\rm BR}\,(\tau\to\mu^-\mu^+\mu^-) & < &  3.2\times 10^{-8} \ , \nonumber \\
{\rm BR}\,(\tau\to\mu^-\gamma) & < & 4.5\times 10^{-8} \ , \nonumber \\
{\rm BR}\,(\tau\to\mu^-\eta) & < & 6.5\times 10^{-8} \ ,  \nonumber \\ 
{\rm and} \ \ 
{\rm BR}\,(\tau\to\mu^-\eta^\prime) & < & 1.3\times 10^{-7} \ .  
\end{eqnarray}
It is clear that these upper limits can easily be saturated by quite 
reasonable values of $\lamp$, for $\widetilde{m} = 300$~GeV. 
Eq.~(\ref{brtau}) also indicates that the upper bound on $\lamp$ scales 
linearly with $\widetilde{m}$.

It is expected \cite{Bona:2007qt} that the Super$B$ factory with 75 
${\rm ab}^{-1}$ luminosity will probe the branching ratios of the 
$\tau^- \to \mu^- \mu^- \mu^+$ and $\tau^- \to \mu^- \eta$ to the level 
of $2 \times 10^{-10}$ and $4 \times 10^{-10}$ respectively. There 
exists the exciting possibility that these LFV processes would be seen 
in such experiments, vindicating the hypothesis of RPV; on the other 
hand, negative results would place stringent bounds on the product 
coupling $\lamp$. At present, however, the best limit on the product 
coupling $\lamp$ arises from BR$(\tau \to \mu\gamma)$, and is given at 
90\% C.L. by
\begin{equation}
\lamp \; \lsim \; 0.3 \left(\frac{\widetilde{m}}{300~{\rm GeV}}\right) \; .
\end{equation}
This is comparable with the individual bounds on $\lambda'_{223}$ and 
$\lambda'_{323}$ mentioned above.

\item \underline{\sl $CP$ violation in the $B_s$--$\bar{B}_s$ system}: 
A very small, but nonvanishing, $\lambda^\prime_{212}~(\sim 0.001)$, 
together with a large $\twtwth(\sim 0.5)$, can generate a large phase in 
$B_s$--$\bar{B}_s$ mixing \cite{Nandi:2006qe}.

\item \underline{\sl Leptonic decays of the $D_s$ meson}: 
The product coupling $\lamp$ can also contribute \cite{Kundu:2008ui} at 
the tree level to the rare decays $D_s \to \ell \nu$ ($\ell\equiv 
\mu,\tau$). At the quark level, the process corresponds to $c \to s \ell 
\nu$. It follows, therefore, that the SM branching ratio would be 
proportional to $G_F^2 |V_{cs}|^2 f_{D_s}^2$.  Once the RPV couplings 
are turned on, the SM formula for the branching ratio in the $D_s \to 
\mu \nu$ channel is modified by replacing
\begin{equation} 
\label{dsmunurpv}
G_F^2 |V_{cs}|^2 \longrightarrow
  \left|G_F V_{cs}^\ast + \frac{\lambda^{\prime 2}_{223}}{4\sqrt{2} {\widetilde
        m}^2} \right|^2 + 
     \left|\frac{\lambda^\prime_{223} \lambda^{\prime\ast}_{323}}
       {4\sqrt{2} {\widetilde m}^2} \right|^2 \, . 
\end{equation} 
For $D_s \to \tau\nu$, we just need to replace $\lambda^\prime_{223} 
\leftrightarrow \lambda^\prime_{323}$ in Eq.~(\ref{dsmunurpv}).  We 
recall at this stage that Monte-Carlo simulations of QCD on the lattice 
\cite{Rosner:2010ak} predict $f_{D_s} = 241~\pm~3$~MeV (HPQCD $+$ UKQCD) 
and $f_{D_s} = 260 \pm 10$ MeV (FNAL $+$ MILC $+$ HPQCD). The 
experimental global average was initially $f_{D_s} = 277 \pm 9$ MeV 
\cite{pdg,:2008sq,:2007ws}. It has now reduced somewhat, viz., $f_{D_s} 
= 257.5 \pm 6.1$ MeV \cite{Rosner:2010ak}, primarily due to the updated 
CLEO result \cite{Naik:2009tk}.

\medskip

A few comments regarding this process are in order. An earlier 
discrepancy between the first lattice result with smaller error bars and 
the initial experimental average had fuelled considerable speculation 
about the possible role of RPV \cite{Kundu:2008ui} or leptoquark 
interactions \cite{Dobrescu:2008er} in the $D_s \to \ell \nu$ channel. 
It was shown \cite{Bhattacharyya:2009hb}, however, that the value of 
$\lamp (\equiv \sqrt{\lprod}) > 0.3$ for $\widetilde{m} = 300$~GeV 
required to explain the $D_s \to \ell \nu$ `anomaly' was too large to be 
consistent with the upper bound $\lamp < 0.3$ from ${\rm Br}~(\tau \to 
\mu\gamma)$. This tension has now disappeared as a result of the new 
experimental average mentioned above, and at present there is no urgency 
to invoke new physics at all in this channel. Nevertheless, the RPV 
couplings $\twtwth$ and $\thtwth$ do stand as potential contributors to 
this mode in case future measurements again create a divide between the 
SM prediction and the experimental result.

\end{itemize} 
\vspace*{-0.2in}
Apart from these cases, where the $\twtwth$ and $\thtwth$ couplings can 
possible generate observable results which differ from the SM, one must 
take into account the possibility that these RPV couplings may be 
constrained by measurements of the anomalous magnetic moment of the 
muon. The maximum possible new physics contribution to this, 
parametrized as $a_\mu \equiv (g_\mu - 2)/2$, is estimated 
\cite{gminus2} as
\begin{equation}
a_\mu^{\rm new} = a_\mu^{\rm exp} - a_\mu^{\rm SM} = (24.6 \pm 8.0) 
\times 10^{-10} \ .
\end{equation} 
In this case, the relevant coupling is $\twtwth$, and the induced 
contribution is given \cite{Bhattacharyya:2009hb} by
\begin{equation} 
\label{amu_numeric}
a_\mu^{\rm RPV} \simeq 1.9 \times 10^{-10} \left(\frac{\twtwth}{0.7}\right)^2 
\left(\frac{300~{\rm GeV}}{\widetilde{m}} \right)^2 \, . 
\end{equation}
Clearly, even if we set $\twtwth \simeq 1$ in Eq.~(\ref{amu_numeric}), 
we cannot saturate $a_\mu^{\rm new}$. This means that the present data 
on the muon anomalous magnetic moment pose no threat to having $\lamp 
\sim 0.3$ with $\widetilde{m} \simeq 300$~GeV.

All in all, with two dominant ($\sim 0.5$) couplings $\twtwth$ and 
$\thtwth$, together with two significantly smaller and optional ones 
($\lambda'_{212} \sim 0.001$ for two-loop neutrino mass generation and 
$\lambda'_{113} < 0.1$ for $B_s$--$\bar{B}_s$ mixing), we can correlate 
several phenomena, real or projected, viz., LFV $\tau$ decays, 
generation of neutrino masses and mixing, leptonic $D_s$ decays, and 
generation of a large phase in $B_s$ mixing, without running afoul of 
constraints from the anomalous magnetic moment of the muon. We emphasize 
that no other viable form of new physics, including the very-similar 
leptoquark interactions, can correlate so many widely-disparate channels 
as our economical choice of only two dominant RPV couplings. With this 
as our motivation, we now turn to the issue of direct searches and seek 
to identify strategies to pinpoint supersymmetric interactions induced 
by these two couplings in the LHC environment.

%%%%%%%%%%%%%%%%%%%%%%%%%%%%%%%%%%%%%%%%%%%%%%%%%%%%%%%%%%%%%%%%%%%%%%%%%%%%%%
\section{Predicting Distinctive Signals at the LHC}

In order to look for a viable process at a hadron collider like the LHC, 
we must take into account the fact that the best kind of signal requires 
a lepton tag and a reasonably large production cross-section. Ideally, 
the new particle production should involve strong interactions to give 
the large production cross-section and its decay should be to a leptonic 
or semi-leptonic channel to provide a good tag. The obvious candidate for 
such a signal is a {\it leptoquark}, i.e. a particle which carries both 
color and lepton number, for it can be produced through QCD processes 
and will inevitably decay into final states containing leptons. As is 
well known, the squarks of an RPV model behave as scalar leptoquarks, 
and hence will be equally good candidates for detection through leptonic 
tags.

Leptonic tags are also possible if the new particles carry a quantum 
number, conserved in strong interactions, which permits pair production 
of heavy particles, but disallows individual particles from decaying 
through strong interactions. This is possible in SUSY, where R parity 
plays the role of the conserved quantum number and the squarks and 
gluinos are produced through their strong interactions at the parton 
level. In the traditional RPC version of SUSY, these squarks and gluinos 
would then decay through cascades to quarks, $W$ and $Z$ bosons and 
eventually to the lightest supersymmetric particle (LSP) \cite{cascades}. 
As is well known, in these models, the LSP is stable and is expected to lead to 
large amounts of missing transverse energy (MET) at a collider such as 
the LHC. This, combined with leptonic tags arising from decays of the 
$W$ and/or the $Z$, produces the trilepton (four leptons) plus jets and 
MET signals which have traditionally been considered 
\cite{SUSYsearch} the most characteristic signatures of 
SUSY\footnote{For alternative sources of such signals at the LHC, 
detailed discussions may be found in Ref.~\cite{JetB}.}.

The previous section has indicated the possibility that having observable 
low-energy signals may demand reasonably large values of two of the possible 
45 RPV couplings, viz., $\lambda'_{223}$ and $\lambda'_{323}$. This will lead 
\cite{BargerHan} to the following interaction vertices:
$$
\lambda'_{223}: \qquad \widetilde{\nu}_\mu-s-b \ , 
                \qquad \widetilde{s}_L-\nu_\mu-b \ ,
                \qquad \widetilde{b}_R-\nu_\mu-s \ ,
                \qquad \widetilde{\mu}_L-c-b \ ,
                \qquad \widetilde{c}_L-\mu-b \ ,
                \qquad \widetilde{b}_R-\mu-c
$$
$$
\lambda'_{323}: \qquad \widetilde{\nu}_\tau-s-b \ , 
                \qquad \widetilde{s}_L-\nu_\tau-b \ ,
                \qquad \widetilde{b}_R-\nu_\tau-s \ ,
                \qquad \widetilde{\tau}_L-c-b \ ,
                \qquad \widetilde{c}_L-\tau-b \ ,
                \qquad \widetilde{b}_R-\tau-c
$$
We now discuss the consequences of having these extra vertices at a 
collider like the LHC, especially the generation of novel processes 
leading to distinctive final states. As R parity remains conserved in 
the gauge sector of the model, the production cross-sections of squarks 
and gluinos will be unchanged. However, the squarks produced either 
directly, or from the decay of a gluino, would now have a competing 
channel to decay directly to a quark and a $\mu^\pm$ or $\tau^\pm$ (or 
to the corresponding neutrinos) through the RPV couplings 
$\lambda'_{223}$ or $\lambda'_{323}$ respectively. This is, in fact, the 
only channel available if the SM is extended by just a leptoquark, 
rather than the full spectrum of RPV supersymmetry. A glance at the 
vertices listed above will show that the most striking processes of this 
form are
\begin{equation}
p + p \toR \widetilde{c}_L {\widetilde{c}}_L^{\,\ast} 
   \to (\mu^+ b) (\mu^- \bar{b})
\qquad\qquad\qquad
p + p \to \widetilde{b}_R \to {\widetilde{b}}_R^{\,\ast} 
   \toR (\mu^- c) (\mu^+ \bar{c})
\end{equation}
and, likewise,
\begin{equation}
p + p \to \widetilde{c}_L \ {\widetilde{c}}_L^{\,\ast} 
   \toR (\tau^+ b) (\tau^- \bar{b})
\qquad\qquad\qquad
p + p \to \widetilde{b}_R \ {\widetilde{b}}_R^{\,\ast} 
   \toR (\tau^- c) (\tau^+ \bar{c})
\end{equation}
The symbol ~($\toR$~)~indicates an R parity-violating decay. Clearly, a 
reconstruction of the invariant mass of the relevant ($\mu^\pm + b$-jet) 
or ($\tau^\pm + b$-jet) will show a resonant peak at the mass of the 
$\widetilde{c}_L$ squark and, likewise, the invariant mass of the 
($\mu^\pm + c$-jet) or ($\tau^\pm + c$--jet) will peak at the mass of 
the $\widetilde{b}_R$ squark. Of these final states, the ones containing 
muons are the most viable since muons are somewhat easier to tag than 
$\tau$ jets in the messy hadronic environment of the LHC. Since this 
work is of exploratory nature, we focus on final states with a muon pair 
accompanied by a pair of hard jets -- with the option of tagging the 
$b$-jets if required -- and of course, with some level of fluctuation in 
jet number due to the usual phenomena of gluon radiation, fragmentation 
and merging of close jets. This is not to say that the signals with 
$\tau$ leptons are to be discounted, for they form a genuine check of 
the model in question, as well as being intrinsically 
detectable\footnote{Especially in view of improved tau-detection 
efficiencies claimed recently \cite{Tonelli}.}. However, the signatures 
and detection strategies for $\tau$ signals will be rather similar to 
those for the muonic signals, except for somewhat lower efficiencies and 
slightly higher backgrounds arising from mistagging of QCD jets as 
$\tau$'s. Since this work is of exploratory nature, we focus on the 
muonic signals and do not consider the $\tau$ signals any further. This 
also means that we discuss signals arising only from the R 
parity-violating coupling $\lambda'_{223}$.
 
We have already noted that a squark $\widetilde{c}_L$ or 
$\widetilde{b}_R$ with R parity-violating couplings closely resembles a 
scalar leptoquark with similar couplings \cite{leptoquark}. Had the only 
new physics component in question consisted of such leptoquarks, then we 
would need to go no further in discussing LHC signals, since these 
leptoquarks would decay solely (i.e. with branching ratio unity) to $\mu 
+ b$ or $\tau + c$ final states. It will then be almost certain, as we 
shall demonstrate in the next section, that leptoquark resonances, if 
they exist and are kinematically accessible, will be seen in the 14~TeV 
run of the LHC.

The situation becomes more complicated --- but also more interesting! 
--- if we consider R parity-violating SUSY, instead of a simplistic 
leptoquark extension of the SM. For then, we also need to take into 
account the entire sparticle spectrum in addition to the squark 
$\widetilde{c}_L$ and $\widetilde{b}_R$, together with their production 
modes and decay channels. Chief among these possibilities is that of 
production and decay of gluinos, which are strongly interacting and will 
be produced in numbers comparable to the squarks. One must also consider 
the charginos and neutralinos, which will couple to the squarks through 
their electroweak gauge couplings and provide competing decay channels 
to the leptoquark-like R parity-violating channels listed above. We must 
also take note of the fact that the gluinos and neutralinos are Majorana 
fermions, and hence can decay into final states with a $\mu^+$ or a 
$\mu^-$ with equal probability --- this gives rise to the possibility of 
same-sign multilepton states which are almost never found in the SM. 
Thus, R parity-violating SUSY permits more exotic signatures than are 
vouchsafed by the mere addition of a leptoquark to the SM.
  
In RPV supersymmetry, as envisaged here, there are two main scenarios, 
depending on the mass of the gluino $\widetilde{g}$. If the gluino 
$\widetilde{g}$ happens to be heavier than the $\widetilde{c}_L$ (and 
$\widetilde{b}_R$), then, in addition to the direct production of 
$\widetilde{c}_L$ (and $\widetilde{b}_R$), these squarks can also arise 
from the cascade decays of gluinos ($\widetilde{g}$), which would be 
pair produced at the LHC. We would then obtain substantial muon-rich 
events from hard processes of the form
\begin{eqnarray}
p + p \to \widetilde{g} \ \widetilde{g}
      \to  (\bar{c} \ \widetilde{c}_L) (\bar{c} \ \widetilde{c}_L)
      \toR \bar{c} \ \bar{c} \ (\mu^+ b) (\mu^+ b)
      &;&
p + p \to \widetilde{g} \ \widetilde{g}
      \to  (\bar{c} \ \widetilde{c}_L) (c \  {\widetilde{c}}_L^{\,\ast})
      \toR c \ \bar{c} \ (\mu^+ b) (\mu^- \bar{b})
\nonumber \\
p + p \to \widetilde{g} \ \widetilde{g}
      \to (c \ {\widetilde{c}}_L^{\,\ast}) (\bar{c} \ \widetilde{c}_L)
      \toR c \ \bar{c} \ (\mu^- \bar{b}) (\mu^+ b) 
      &;&
p + p \to \widetilde{g} \ \widetilde{g} 
      \to (c \  {\widetilde{c}}_L^{\,\ast}) (c \  {\widetilde{c}}_L^{\,\ast})
      \toR c \ c \ (\mu^- \bar{b}) (\mu^- \bar{b})
\nonumber \\
\end{eqnarray}
i.e. final states with a muon pair of like or unlike sign, typically 
accompanied by {\it four} hard jets, subject to jet-merging and other 
strong interaction effects, as listed above. Gluinos, being Majorana 
fermions, can decay with equal probability into $\widetilde{c}_L$ or 
$\widetilde{c}^{\,\ast}_L$ states, which accounts for the like- and 
unlike-sign muons in the final state. To every process listed above, 
there corresponds a similar process with the $\widetilde{b}_R$ squarks: 
these are not listed in the interests of brevity.

If, on the other hand, the gluino is lighter than the $\widetilde{c}_L$ 
and $\widetilde{b}_R$, these squarks will tend to decay dominantly 
through the strong interaction to $\widetilde{c}_L \to \widetilde{g} + 
c$ and $\widetilde{b}_R \to \widetilde{g} + b$. The competing RPV 
decays will be suppressed, and hence, we will get a somewhat smaller 
multi-lepton signal of the form envisaged above. However, we note that 
the desire to have observable flavor physics effects prompts us to take 
\begin{equation}
\lambda'_{223} \simeq 
 0.3 \left(\frac{M(\widetilde{c}_L)}{300~{\rm GeV}} \right) =  
\frac{M(\widetilde{c}_L)}{1~{\rm TeV}}
\label{eqn:scaling}
\end{equation}
for slightly larger values of $M(\widetilde{c}_L)$. Thus, the branching ratio 
controlled by $\lambda'_{223}$ grows larger for heavy $\widetilde{c}_L$, and 
can remain substantial even when the gluino is lighter than the 
$\widetilde{c}_L$. This
is illustrated in Figure~\ref{fig:clbr} for both the cases -- where the 
gluino is lighter than the $\widetilde{c}_L$ squark and vice versa.
We further note that if $M(\widetilde{c}_L) = 1.5$~TeV, Eqn.~(\ref{eqn:scaling})
$\lambda'_{223} \simeq 1.5$, which is close to the end of the perturbative
regime. Our discussion, therefore, is restricted only to values of 
$M(\widetilde{c}_L) \leq 1.5$~TeV. 

%%%%%%%%%%%%%%%%%%%%%%%%%%%%%%%%%%%%%%%%%%%%%%%%%%%%%%%%%%%%%%%%%%%%%%%%%%%%%
\begin{figure}[htb]
\setcounter{figure}{0}
\centerline{ \epsfxsize= 5.0 in \epsfysize= 3.0in \epsfbox{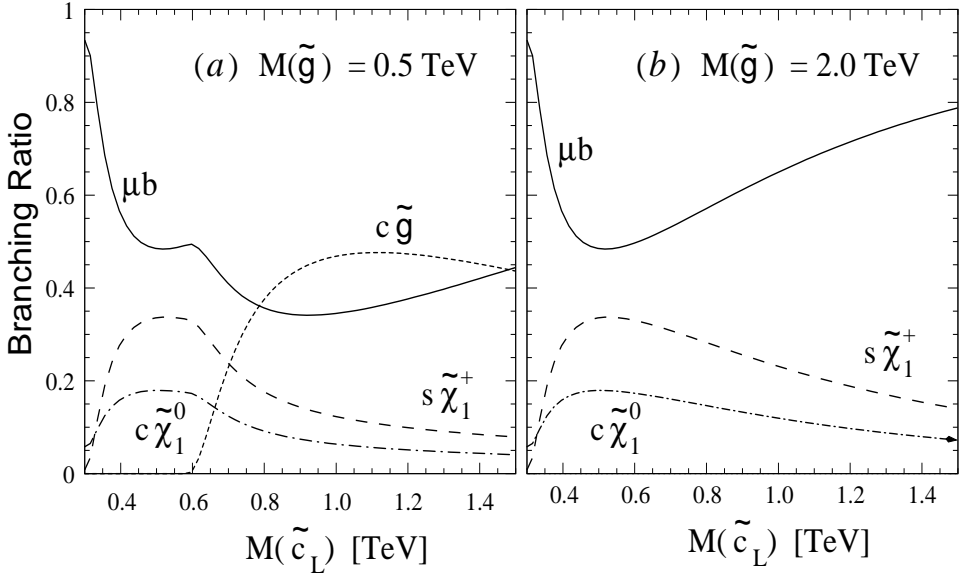} }
\vskip -1pt
\def\baselinestretch{0.8}
\caption{{\footnotesize Illustrating the variation with the 
$\widetilde{c}_L$ mass of branching ratios for the processes
$\widetilde{c}_L \to \mu b$, $\widetilde{c}_L \to c\widetilde{g}$,
$\widetilde{c}_L \to s\widetilde{\chi}_1^+$ and $\widetilde{c}_L
\to c\widetilde{\chi}_1^0$ for the cases when ($a$) the gluino mass
is 500~GeV and ($b$) the gluino mass is 2~TeV. For this illustrative
plot, the other relevant MSSM parameters have been taken as $M_1 = 100$~GeV, 
$M_2 = 300$~GeV, $\mu = 1$~TeV while all squark (other than 
$\widetilde{c}_L$) and slepton masses are set to 1.5~TeV.}}
\def\baselinestretch{1.2}
\label{fig:clbr}
\end{figure}
%%%%%%%%%%%%%%%%%%%%%%%%%%%%%%%%%%%%%%%%%%%%%%%%%%%%%%%%%%%%%%%%%%%%%%%%%%%%% 
\vskip -5pt

A glance at Figure~\ref{fig:clbr}($a$) will show that the direct RPV
decay $\widetilde{c}_L \to \mu b$ always dominates the chargino and
neutralino channels, and is competitive with the gluino channel when that 
opens up. In the right panel, Figure~\ref{fig:clbr}($b$) shows that 
$\widetilde{c}_L \to \mu b$ remains the dominant channel everywhere 
if the gluino channel is kinematically disallowed. This is also the case
where gluino production followed by cascade decays will enhance the 
$\mu b$ signal. However, the LHC cross-sections will start 
falling for squark and gluino masses above a TeV. 

If we now turn to the decay modes of the $\widetilde{c}_L$ through a 
gluino or a chargino or a neutralino, we note that new decay channels 
are opened here too because of R parity violation. Thus, a gluino or a 
neutralino can undergo a three-body decay through an off-shell 
$\widetilde{c}_L$ or $\widetilde{c}_L^{\,\ast}$ as
\begin{eqnarray}
\widetilde{g} ~ ~ {\rm or} ~ ~ \widetilde{\chi}_{i}^0 
& \to & \bar{c} \ \widetilde{c}_L \toR \bar{c} \ (\mu^+ b) \nonumber \\
& \hookrightarrow & c \ {\widetilde{c}}_L^{~*} \toR c \ (\mu^- \bar{b})
\end{eqnarray}
for all the neutralinos $i = 1,2,3,4$ including the lightest one, which is 
usually the LSP. Similarly the charginos can decay as
\begin{equation}
\widetilde{\chi}_{i}^+ \to \bar{s} \ \widetilde{c}_L \toR \bar{s} \ (\mu^+ b)
\qquad\qquad\qquad
\widetilde{\chi}_{i}^- \to s \ {\widetilde{c}}_L^{\,\ast} \toR s \ 
(\mu^- \bar{b})
\end{equation}
where $i = 1,2$. In such cases the final state will again contain a muon 
pair and a pair of $b$ jets among other jets, but in this case we can no 
longer expect the $\mu - b$ invariant mass to peak at the 
$\widetilde{c}_L$ mass.

It is clear, therefore, that in a supersymmetric model with RPV, the 
possible decays of the $\widetilde{c}_L$ can occur through many 
channels, each of them producing a final state with a muon pair and two 
$b$ quarks, with other quarks in the final state as well. At the parton 
level, a partial list is given below.
$$
\begin{array}{lllllllllllr}
\widetilde{c}_L & \toR & b ~\mu^+ &&&&&&&&& (16) \\ \\

& \hookrightarrow &s ~\widetilde{\chi}^+_{i} 
& \toR & s ~\bar{s} ~b ~\mu^+  & &(i = 1,2) &&&&&  \\
&  & & \hookrightarrow & s ~W^+ ~\widetilde{\chi}_1^0 
& \to & s ~\ell^+ \nu_\ell ~\widetilde{\chi}_1^0 & \toR 
& s ~\ell^+ \nu_\ell ~(\bar{c} ~b ~\mu^+)  & {\rm or} 
& s ~\ell^+ \nu_\ell ~(c ~\bar{b} ~\mu^-) & \\ 
&  & &  & 
& \hookrightarrow & s ~q ~\bar{q}'~\widetilde{\chi}_1^0 & \toR  
& s ~q ~\bar{q}' ~(\bar{c} ~b ~\mu^+)  & {\rm or} 
& s ~q ~\bar{q}' ~(c ~\bar{b} ~\mu^-) & \\

&  & & \hookrightarrow & s ~\ell^+ ~\widetilde{\nu}_\ell 
& \to & s ~\ell^+ ~\nu_\ell ~\widetilde{\chi}_1^0  & \toR  
& s ~\ell^+ \nu_\ell ~(\bar{c} ~b ~\mu^+) & {\rm or} 
& s ~\ell^+ \nu_\ell ~(c ~\bar{b} ~\mu^-) &\\

&  & & \hookrightarrow & s ~\nu_\ell ~\widetilde{\ell}^+ 
& \to & s ~\nu_\ell ~\ell^+ ~\widetilde{\chi}_1^0  & \toR  
& s ~\ell^+ \nu_\ell ~(\bar{c} ~b ~\mu^+) & {\rm or} 
& s ~\ell^+ \nu_\ell ~(c ~\bar{b} ~\mu^-) &\\ \\

& \hookrightarrow & c ~\widetilde{\chi}_1^0 
& \toR & c ~\bar{c} ~b ~\mu^+ & {\rm or} 
& \bar{c} ~c ~\bar{b} ~\mu^- &&&&& \\ \\

& \hookrightarrow & c ~\widetilde{\chi}_j^0 
& \toR & c ~\bar{c} ~b ~\mu^+ & {\rm or} 
& \bar{c} ~c ~\bar{b} ~\mu^- &&(j = 2,3,4)&&&  \\ 

& & & \hookrightarrow & c ~Z^0 ~\widetilde{\chi}_1^0 & \to 
& c ~\ell^+\ell^- ~\widetilde{\chi}_1^0 & \toR
& c ~\ell^+\ell^- ~(\bar{c} ~b ~\mu^+) & {\rm or} 
& c ~\ell^+\ell^- ~(c ~\bar{b} ~\mu^-) & \\ 

& & & & & \hookrightarrow & c ~q ~\bar{q} ~\widetilde{\chi}_1^0 & \toR
& c ~q ~\bar{q} ~(\bar{c} ~b ~\mu^+) & {\rm or} 
& c ~q ~\bar{q} ~(c ~\bar{b} ~\mu^-) & \\  
 
& & & \hookrightarrow & c ~h^0 ~\widetilde{\chi}_1^0
& \to & c ~b ~\bar{b} ~\widetilde{\chi}_1^0 & \toR
& c ~b ~\bar{b} ~(\bar{c} ~b ~\mu^+) & {\rm or} 
& c ~b ~\bar{b} ~(c ~\bar{b} ~\mu^-) & \\ 

& & & \hookrightarrow & c ~\ell^\pm ~\widetilde{\ell}^\mp 
& \to & c ~\ell^+\ell^- ~\widetilde{\chi}_1^0  &\toR  
& c ~\ell^+\ell^- ~(\bar{c} ~b ~\mu^+) & {\rm or} 
& c ~\ell^+\ell^- ~(c ~\bar{b} ~\mu^-) &\\

& & & \hookrightarrow & c ~\bar{\nu}_\ell ~\widetilde{\nu}_\ell 
& \to & c ~\bar{\nu}_\ell ~\nu_\ell ~\widetilde{\chi}_1^0  &\toR  
& c ~\bar{\nu}_\ell ~\nu_\ell ~(\bar{c} ~b ~\mu^+) & {\rm or} 
& c ~\bar{\nu}_\ell ~\nu_\ell ~(c ~\bar{b} ~\mu^-) &\\

\end{array}
$$
In the above, $\ell = e, \mu$, while $q\bar{q}$ refers to any pair of 
quarks $q = u,d, s, c, b$, and $q\bar{q}' = u\bar{d}$ or $c\bar{s}$.  
The use of alternatives marked `or' corresponds to the Majorana nature 
of the neutralinos. Even if we assume that the gluino is heavier that 
the $\widetilde{c}_L$, this list is illustrative, but not exhaustive, 
since there is considerable dependence on the mass spectrum of the 
sparticles, i.e. on the choice of supersymmetry-breaking parameters. For 
example, in the above decay chains, it has been assumed that all the 
squarks are heavier than the charginos and neutralinos, but if there is 
a light squark, additional decay channels of these particles will open 
up. Similarly, it has been assumed that the sleptons and sneutrinos are 
lighter than the $\widetilde{\chi}_{1}^\pm$ and $\widetilde{\chi}_2^0$ ---
but heavier, of course, than the LSP $\widetilde{\chi}_1^0$.  
If this is violated, some of 
the above decay channels will be kinematically disallowed. Moreover, 
kinematics permitting, we can also have decay modes involving the 
heavier $H^0$, $A^0$ and $H^\pm$ states. Again, if the $\widetilde{c}_L$ 
can decay into a lighter gluino $\widetilde{g}$ many more channels are 
opened up. All these complications are, in fact, taken care of in our 
numerical studies, but it is not our purpose to display them here in an 
infinity of detail. Rather, the purpose of exhibiting the above list is 
to point out the following interesting facts, viz.
\vspace*{-0.2in}
\begin{itemize}

\item The cascade decays occurring through RPC couplings will end up 
in a $\widetilde{\chi}_1^0$ which will decay with equal probability
into $c\bar{b}\mu^-$ and $\bar{c}b\mu^+$ final states, shown enclosed between
parentheses in the last column of Eqn.~(16). Given that the direct
$\mu b$ decay is always at the level of 40\% or more (see Figure~\ref{fig:clbr})
it follows that the decay products of a $\widetilde{c}_L$ will contain a 
muon --- of indeterminate sign --- and a $b$ quark or antiquark in 
at least 70\% of the events, so long as there is a $\lambda'_{223}$ interaction.

\item When $\widetilde{c}_L$'s are pair-produced and if they decay 
directly through the $\lambda'_{223}$ coupling, the final state will 
contain a muon pair and two $b$ jets. In this case, matching resonances 
will be seen in 2 of the 4 possible $\mu b$ invariant masses. This is 
the only signal if the new physics consists of a leptoquark rather than 
squarks of RPV supersymmetry.

\item If the $\widetilde{c}_L$ pair originates from the decays of a 
gluino pair, and then decays to a muon pair and two $b$ jets, there is a 
possibility of same-sign muon pairs because of the Majorana nature of 
the gluino. Once again, there will be matching resonances in two $\mu b$ 
invariant masses.

\item If $\widetilde{c}_L$'s are pair-produced, and decay mainly through 
RPC channels (e.g. through a gluino, chargino or neutralino), there will 
be no resonances in $\mu b$ invariant masses, but one can have same-sign 
muon pairs arising from the neutralino LSP decays. An excess of muon 
pairs in the overall tally of same-sign lepton pairs is indicative of a 
$\lambda'_{223}$ coupling, as against a RPC scenario, where lepton 
flavor universality is maintained.

\item If the first step in the cascade decay of a $\widetilde{c}_L$ is 
through a chargino, decaying semileptonically, and the decay chain ends 
in a neutralino LSP decaying through RPV couplings, there exists a 
possibility of having same-sign leptons triads in the final state. These 
are so difficult to produce in other models that they practically 
constitute a `smoking gun' signal for R parity violation \cite{HOD}.

\item If a $\widetilde{c}_L$ pair arises in the decay of a gluino pair, 
and each $\widetilde{c}_L$ undergoes a decay beginning with a chargino 
and ending in a decaying neutralino LSP, one can even have spectacular 
events with lepton quartets of the same sign. Rare as such events will 
be, they will have no background(s) worth speaking of \cite{HOD}.

\end{itemize}
\vspace*{-0.2in}
\noindent We now conduct a collider study for the $\lambda'_{223}$ 
coupling by following the agenda set by the list of observations made 
above. The following section discusses the issue of resonances in $\mu 
b$ invariant masses in the different cases explained above.

%%%%%%%%%%%%%%%%%%%%%%%%%%%%%%%%%%%%%%%%%%%%%%%%%%%%%%%%%%%%%%%%%%%%%%%%%%%%%%
\section{Scalar Resonances in Semileptonic Final States}

In order to study $\widetilde{c}_L$ production and decay numerically we 
perform a Monte Carlo simulation of the relevant processes using the 
well-known event generator {\sc pythia}, which has an option to include 
RPV couplings at will, in addition to algorithms for fragmentation and 
jet formation as well as initial- and final-state radiation of multiple 
soft gluons \cite{Pythia}. We also use the {\sc pycell} algorithm 
inbuilt in the software to identify and segregate jets. While it is 
known that {\sc pycell} is not the best of all the available jet 
identification algorithms, it is good enough for an exploratory study 
such as the present one. For the RPC part of the SUSY spectrum, we use 
the software {\sc SuSpect} \cite{Suspect}. Armed with these tools, we 
can now commence on a study of the RPV model in question and its 
signals.

Our analysis is carried out in two stages, as detailed below.
\vspace*{-0.2in}
\begin{itemize}
\item We first consider the pair-production of $\widetilde{c}_L$'s
where each $\widetilde{c}_L$ decays to $\mu b$ with unit 
branching ratio. This corresponds to the cases when ($i$) the 
SM is merely extended by the addition of a leptoquark\footnote{Actually 
two leptoquarks $\widetilde{c}_L$ and $\widetilde{b}_R$, to fit
the low-energy requirements.} decaying wholly to $\mu b$ final states,
and ($b$) the $\widetilde{c}_L$ is the LSP in an RPV version of the 
MSSM where all other SUSY particles are too heavy to have appreciable
cross-sections. This is essentially the study presented in the current 
section, though we shall eventually consider some other scenarios. 
\item We then take up the case when the $\widetilde{c}_L$ is embedded 
in a RPV model with chargino(s), neutralino(s), and maybe
gluinos, of comparable masses and cross-sections. This modifies the previous search in two ways, viz. ($i$)~by 
reducing the branching ratio for $\widetilde{c}_L \to \mu b$ from unity, and 
($ii$) by introducing new $\widetilde{c}_L$ production modes through
SUSY cascades. In this section we shall consider only the effects of the 
change in branching ratio. A detailed simulation of this scenario is carried
out and described in the next section. 
\end{itemize}
\vspace*{-0.2in}

We first take up the case of a leptoquark, or a $\widetilde{c}_L$ LSP with 
other SUSY particles being very heavy (i.e. contributing neither to the
signal or the background). We shall continue, however, to denote the particle 
as a $\widetilde{c}_L$. Obviously, the cleanest signal for a 
leptoquark/squark $\widetilde{c}_L$
with a $\lambda'_{223}$ coupling would arise when we 
observe two $\mu + b$ pairs, each with invariant mass peaking around the mass of the $\widetilde{c}_L$. Since muons are easily identified at the LHC 
detectors, the viability of this signal is really controlled by the 
efficiency with which $b$ jets can be tagged. A detailed study 
by the ATLAS Collaboration \cite{ATLAS:b_tag} shows that the efficiency 
of $b$ tagging is in the neighbourhood of 40\% for $b$ jets with $p_T$ 
less than about 300~GeV, but for higher $p_T$ jets, the efficiency 
declines sharply. It follows that $b$ tagging is a useful option only if 
the candidate jets have $p_T \; \lsim \ 300$~GeV. When the $p_T$ of the 
$b$-parton is significantly higher than this, we will have to treat 
$b$-jets on the same footing as light quark jets. We must, therefore, 
concern ourselves with the question: what are the final states of 
interest when a pair of $\widetilde{c}_L$ resonances decays into a muon 
pair and a pair of $b$-jets, which may or may not be tagged? A little 
reflection will show that there are four important possibilities, as 
detailed below.
\vspace*{-0.2in}
\begin{enumerate}

\item \underline{\it The final state contains two muons and {\it 
exactly} two jets, both of which are tagged as $b$ jets.} \\ If we label 
the muons $\mu_1$ and $\mu_2$ by hardness (i.e. by the $p_T$ value) and 
similarly label the $b$-jets as $b_1$ and $b_2$ by hardness, we can now 
construct four invariant masses, viz. $M(\mu_1 b_1)$, $M(\mu_1 b_2)$, 
$M(\mu_2 b_1)$ and $M(\mu_2 b_2)$. Of these, one pair of invariant 
masses will correspond to the pair of $\widetilde{c}_L$ resonances, and 
hence will have similar values. We, therefore, construct the ratios
\begin{equation}
\setcounter{equation}{17}
R_1 = \frac{M(\mu_1 b_1)}{M(\mu_2 b_2)} \qquad {\rm and} \qquad
R_2 = \frac{M(\mu_1 b_2)}{M(\mu_2 b_1)}
\end{equation}
Of these, one ratio will be close to unity, since it will correspond to 
the correct choice of muon and $b$-jets, while the other will have some 
random value. We, therefore denote the correct ratio simply as $R$ and 
impose a kinematic criterion requiring $R$ to be close to unity. It is 
important to note that in tagging two $b$-jets with an efficiency around 
40\% each, we retain only 16\% of the original events and hence, this 
signal may be rather small.

\item \underline{\it The final state contains two muons and 2 -- 3 jets, 
of which two are tagged as $b$-jets.} \\ This will correspond to the 
same process as above, but will include the cases where ($i$) one of the 
$b$-jets fragments into two, or ($ii$) there is an extra jet due to a 
gluon radiation from either the initial or the final state. Such cases 
would be excluded from the previous tag. Only the jets tagged as 
$b$-jets will be used to construct invariant masses and ratios thereof. 
This signal will thus resemble the previous one, but will be 
significantly enhanced.

\item \underline{\it The final state contains two muons and {\it 
exactly} two jets.} \\ No $b$-tagging is employed in this case, and 
hence, the number of events will be much larger. As before, if we label 
the muons $\mu_1$ and $\mu_2$ by and the jets as $J_1$ and $J_2$ by 
hardness, we can now construct four invariant masses, viz. $M(\mu_1 
J_1)$, $M(\mu_1 J_2)$, $M(\mu_2 J_1)$ and $M(\mu_2 J_2)$. Once again, we 
can construct ratios
\begin{equation}
R_1 = \frac{M(\mu_1 J_1)}{M(\mu_2 J_2)} \qquad {\rm and} \qquad
R_2 = \frac{M(\mu_1 J_2)}{M(\mu_2 J_1)}
\end{equation}
and demand that one of them lie close to unity. This signal will be at 
least an order of magnitude higher than the similar case when we tag the 
$b$ jets, for reasons explained above. However, it is no longer directly 
related to the coupling $\lambda'_{223}$ but can be initiated by any of 
the $\lambda_{22i}$, where $i = 1,2,3$.

\item \underline{\it The final state contains two muons and 2 -- 3 
jets.} \\ This is an inclusive case of the previous signal in which we 
allow an extra jet which may have been generated by QCD effects. Once 
again, no $b$-tagging is employed, and hence, if there are three jets, 
ordered $J_1$, $J_2$ and $J_3$ by hardness, one has to construct six 
invariant masses $M(\mu_1 J_1)$, $M(\mu_1 J_2)$, $M(\mu_1 J_3)$, 
$M(\mu_2 J_1)$, $M(\mu_2 J_2)$, and $M(\mu_2 J_3)$, and six ratios
\begin{eqnarray}
R_1 = \frac{M(\mu_1 J_1)}{M(\mu_2 J_2)} \qquad 
R_2 = \frac{M(\mu_1 J_1)}{M(\mu_2 J_3)} \qquad 
R_3 = \frac{M(\mu_1 J_2)}{M(\mu_2 J_1)} \nonumber \\ 
R_4 = \frac{M(\mu_1 J_2)}{M(\mu_2 J_3)} \qquad 
R_5 = \frac{M(\mu_1 J_3)}{M(\mu_2 J_1)} \qquad 
R_6 = \frac{M(\mu_1 J_3)}{M(\mu_2 J_2)} \nonumber
\end{eqnarray}
of which one may be expected to peak near unity. Obviously this signal 
yields the most events, since there is no suppression due to 
$b$-tagging, and most of the QCD NLO effects are taken into account.

To avoid combinatorial backgrounds, this criterion may be also imposed 
in a cruder form, by assuming that $J_3$, the softest of the jets will 
mostly arise from QCD effects rather that from a $\widetilde{c}_L$ 
decay. In this case, we can just construct $R_1$ and $R_2$ using the 
jets $J_1$ and $J_2$ and require one of them to be close to unity. In 
our numerical analysis, we have tried both approaches and found them to 
yield comparable efficiencies.
 
\end{enumerate}
\vspace*{-0.2in}
The four kinds of signal described 
above will be discernible only if the SM backgrounds are sufficiently small.  
The first and most obvious of these will arise from $t\bar{t}$ 
production, which has an NLO cross-section close to a nanobarn at the 14~TeV LHC. This is huge 
compared with the production cross-section for $\widetilde{c}_L$ 
squarks, which ranges from a few femtobarns to a few picobarns, 
depending on the squark mass. Moreover, the semileptonic decays of a top 
quark (or antiquark) will almost always contain a $b$-jet, and will have 
a 10.6\% branching ratio to muons. Thus, there will be no dearth of 
final states containing a muon pair and a pair of $b$-jets in the decay 
products following a $t\bar{t}$ event, and it follows that it is a 
challenge to devise kinematic criteria which will restrict this 
background to manageable levels. Other backgrounds arise from the 
pair-production of $W^+W^-$, $W^\pm Z$ and $ZZ$ accompanied by varying 
numbers of QCD jets. In particular, we could have, for example, $ZZ \to 
(\mu^+\mu^-)(b\bar{b})$ or $WZ \to \mu + \not{\!\!E}_T + {\rm jets}$ or 
$W^+W^-$~+~jets $\to \mu^+\mu^- + \not{\!\!E}_T$ + jets. Of course, our 
signal does not have substantial MET, but there will always be 
configurations with low MET among the background events, and even though 
these are rare, the cross-section for vector boson pair production (tens 
of picobarns) is enough to make their contribution significant. If a 
single $Z$, accompanied by a pair of $b$-jets, decays to $\mu^+\mu^-$, 
this can also be a sizable background.

\begin{figure}[htb]
\centerline{ \epsfxsize= 4.0 in \epsfysize= 3.75in \epsfbox{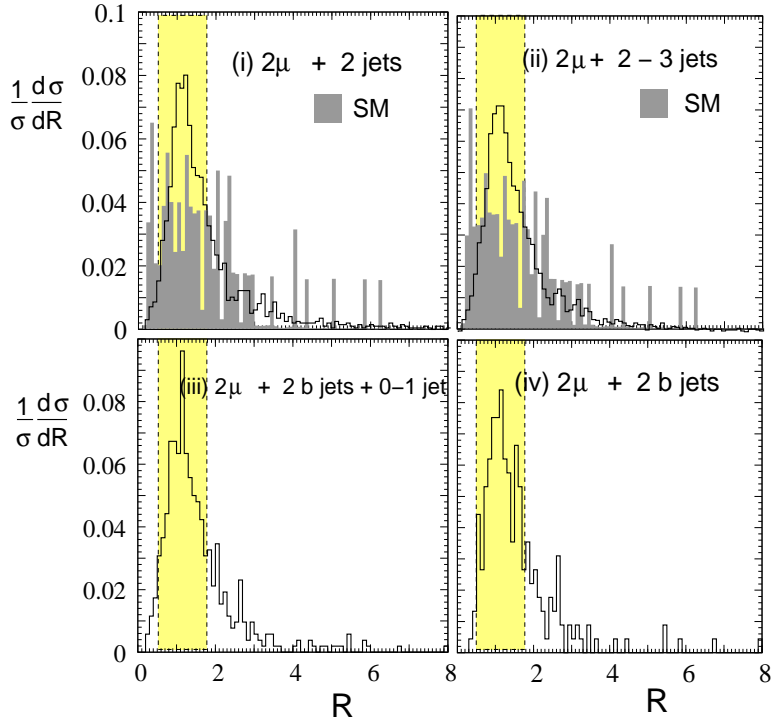} }
\vskip -1pt
\def\baselinestretch{0.8}
\caption{{\footnotesize Normalized distribution in $R$, the ratio of 
invariant masses for the four cases where the final state contains two 
muons and ($i$) exactly two jets, ($ii$) up to three jets ($iii$) two 
$b$-jets and up to one extra jet, and ($iv$) exactly two $b$ jets. 
The $\widetilde{c}_L$ mass is taken as 600~GeV. The 
shaded histogram in the upper panels represents the SM background. In 
the lower panels, the SM background is practically absent. The (yellow) 
shaded block between the broken lines indicates the cut on $R$.}}
\def\baselinestretch{1.2}
\label{fig:Ratio}
\end{figure}

It is clear, therefore, that the mere appearance of a muon pair 
accompanied by jets and little or no missing energy is not at all a 
clear signal for the particle in question. There are large irreducible 
SM backgrounds which must be taken into consideration. In principle, 
however, we have a panacea for all these backgrounds using our kinematic 
criterion imposed on the ratios $R_1$ and $R_2$ (see above). 
These will, in general, be close to unity only when a pair of 
$\widetilde{c}_L$ particles is produced and each of these then decay to 
a $\mu + b$ final 
state. For the $t\bar{t}$ and vector boson backgrounds, there is no 
\'a priori reason, except 
statistical fluctuations, for this ratio to lie in the neighbourhood of 
unity. This is illustrated in Figure~\ref{fig:Ratio}, where we have 
plotted the signal and background distributions in the $R$ which is 
closest to unity for the four classes of signal described above.

In order to generate Figure~\ref{fig:Ratio}, we have made use of the 
following kinematic cuts and efficiency estimates.
\vspace*{-0.2in}
\begin{enumerate}

\item The $b$-tagging efficiency is assumed to be $\epsilon_b = 0.4$ 
uniformly, for $p_T(b) \leq 300$~GeV. This is not a very bad 
approximation, since the variation in $\epsilon_b$ is rather small in 
the kinematic range in which the bulk of the $b$-jets are found in our 
analysis \cite{ATLAS:b_tag}. Similarly, we have assumed that the 
probability of mistagging a $c$-quark jet as a $b$ jet is 10\% and the 
probability of mistagging a light quark/gluon jet as a $b$-quark is 
0.5\%.

\item For the leptons (muons and electrons) we have imposed the 
following isolation cut: the hadronic energy deposit in a cone of radius 
$\Delta R = 0.2$ around the lepton direction should be less than 10~GeV.

\item For the $2\mu +$~jets signal, we impose $\not{\!\!E}_T < 150~{\rm 
GeV}$ as well as
$$
\frac{\not{\!\!E}_T}{M_{\rm eff}} < 0.1 \ .
$$
If we label the muons as $\mu_1, \mu_2$ and the jets as J$_1$, J$_2$ 
according to their transverse momenta $p_T$, then we also impose
$$
p_T(\mu_1), \; p_T({\rm J}_1) > 100~{\rm GeV} \; ,
\qquad\qquad
p_T(\mu_2), \; p_T({\rm J}_2) > 50~{\rm GeV} \; ,
$$ 
together with
$$
M(\mu\mu), M(b\bar{b}) > 150~{\rm GeV} \, .
$$

\item Finally we impose the kinematic criterion that if $R$ be the ratio 
of invariant masses which lies closer to unity, then $ 0.5 < R < 1.8$.

\end{enumerate}

%%%%%%%%%%%%%%%%%%%%%%%%%%%%%%%%%%%%%%%%%%%%%%%%%%%%%%%%%%%%%%%%%%%%%%%%%%%%%
%%%%%%%%%%%%%%%%%%%%%%%%%%%%%%%%%%%%%%%%%%%%%%%%%%%%%%%%%%%%%%%%%%%%%%%%%%%%% 
\vskip -5pt

The four panels in Figure~\ref{fig:Ratio} correspond to the four kinds 
of signal marked, i.e. running clockwise from the upper left panel: muon 
pair + 2 jets, muon pair + 2--3 jets, muon pair + 2 $b$ jets and muon 
pair + 2 $b$ jets + 0--1 jet. Normalized distributions have been shown 
in each case. The unfilled histograms indicate the signal and 
the dark grey shaded histogram represents the total SM background.
To generate these plots, the $\widetilde{c}_L$ mass has been 
taken as 600 GeV, but the distribution is not sensitive to reasonable 
variations in this parameter.  No 
background is indicated in the lower two panels because these backgrounds lie 
below the level of a single event in a bin, assuming $\sqrt{s} =
14$~TeV and an integrated luminosity of 10~fb$^{-1}$. The broken lines indicate the 
range of $R$ which we have considered as permissible for the signal, 
i.e. $0.5 < R < 1.8$.

A glance at Figure~\ref{fig:Ratio} will show that the signal is not as 
sharply peaked around unity as one would have expected from a resonant 
particle with a decay width of a few GeV at most. The reason is not far 
to seek: the reconstruction of the final state jets is fraught with 
uncertainties, arising from, for example, final state radiation, jet 
splitting/merging and/or undetected momentum in the form of soft jets or 
neutrinos. This results in a smearing-out of the peak, generating a long 
tail, which is apparent in all four panels of the Figure. However, most 
of the signal events are contained in the chosen range, viz. $R = 0.5 - 
1.8$, as can be seen from the graph. It is somewhat unfortunate that 
this is also the area where a significant portion of the background lies, 
though that forms a much broader distribution. In fact, the cut $0.5 < R < 1.8$ 
removes about half of the SM background while leaving the leptoquark/RPV 
signal comparatively unscathed (even though a small tail towards large values 
of $R$ does get affected).

One may question the rationale for imposing this criterion for the cases 
with $b$ tagging, represented by the lower pair of boxes in 
Figure~\ref{fig:Ratio}, given that the background is vanishingly small. 
The answer is that this criterion isolates events which are consistent with 
the hypothesis of a {\it pair} of heavy particles being produced and 
each then decaying into a muon and a $b$-jet. Events which satisfy this 
cut, therefore, may be considered candidates for a leptoquark/RPV signal. Thus the kinematic criterion on ratios, though it does not prove as powerful 
as one may have 
wished, certainly enriches the signal content in the selected sample of 
events.

%%%%%%%%%%%%%%%%%%%%%%%%%%%%%%%%%%%%%%%%%%%%%%%%%%%%%%%%%%%%%%%%%%%%%%%%%%%%%
\begin{figure}[htb]
\centerline{ \epsfxsize= 2.8 in \epsfysize= 2.8 in \epsfbox{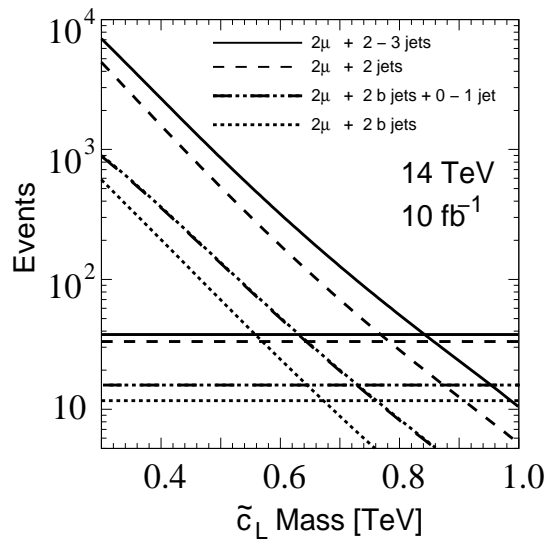} }
\vskip -1pt
\def\baselinestretch{0.8}
\caption{{\footnotesize\ Illustrating the variation in total 
cross-sections as the mass of the leptoquark/squark $\widetilde{c}_L$ is varied, for the 
four kinds of signal considered in Figure~\ref{fig:Ratio}, after implementation 
of the ratio criterion. 
The horizontal lines correspond to the 
SM background. We take $B(\widetilde{c}_L \to \mu^+ b) = 1$ for the 
signal, as expected for a leptoquark.}}
\def\baselinestretch{1.2}
\label{fig:CrossSection}
\end{figure}
%%%%%%%%%%%%%%%%%%%%%%%%%%%%%%%%%%%%%%%%%%%%%%%%%%%%%%%%%%%%%%%%%%%%%%%%%%%%% 
\vskip -5pt

In Figure~\ref{fig:CrossSection} we show how the total cross-section for 
these four kinds of final state varies as the mass of the  
leptoquark/squark $\widetilde{c}_L$ increases from 300~GeV to 1~TeV, 
which is the LHC range of interest. These graphs make it clear that $b$-
tagging is not a very good 
strategy, especially when the $\widetilde{c}_L$ mass increases above 
$\sim 550$~GeV. At the full energy of 14~TeV and an integrated 
luminosity of 10~fb$^{-1}$, the fact that tagging two $b$ jets leads to 
a low efficiency factor of around 0.16 can be tolerated, since the 
statistics are large. However, in the early runs, the luminosity 
collection will be low and hence $b$-tagging will be a serious issue. In 
any case, a glance at Figure~\ref{fig:CrossSection} shows that, as a 
search for new physics, the signals {\it without} $b$-tagging have a much 
better kinematic reach. For example, we can have a $5\sigma$ signal all 
the way up to $M(\widetilde{c}_L) = 720 (780)$~GeV for the signal with a 
muon pair and two jets (up to three jets). For even higher masses of the 
$\widetilde{c}_L$, it is no longer worth considering the total 
cross-section, but we should look at the differential cross-section, 
where there will be resonant peaks in the muon-jet invariant mass 
construction. 

%%%%%%%%%%%%%%%%%%%%%%%%%%%%%%%%%%%%%%%%%%%%%%%%%%%%%%%%%%%%%%%%%%%%%%%%%%%%%
\begin{figure}[htb]
\centerline{ \epsfxsize= 6.5 in \epsfysize= 2.5 in \epsfbox{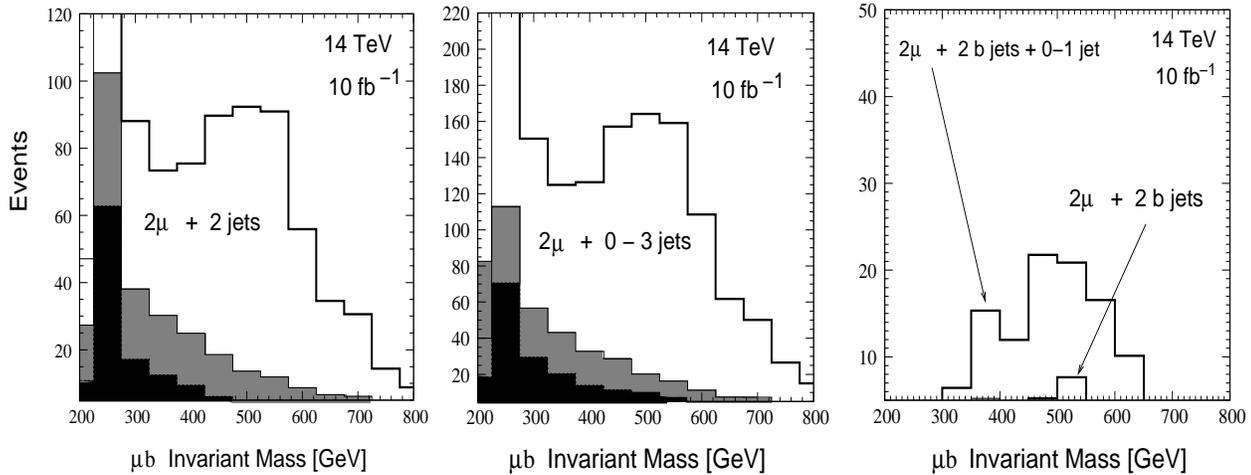} }
\vskip -1pt
\def\baselinestretch{0.8}
\caption{{\footnotesize $\mu b$ invariant mass distribution for the same 
leptoquark/squark signals as in Figure~\ref{fig:Ratio} with 
$B(\widetilde{c}_L \to \mu^+ b) = 1$ as before. In each panel, the 
unshaded histogram indicates the signal added to the SM background, which is 
indicated by the dark-shaded histogram on the left and centre. The 
light-shaded histogram represents the $5\sigma$ fluctuation in the SM 
background. Note that the lower end of the plot corresponds to 5 events: 
in the panel on the right the SM backgrounds never rise above this 
minimum.}}
\def\baselinestretch{1.2}
\label{fig:InvMass}
\end{figure}
%%%%%%%%%%%%%%%%%%%%%%%%%%%%%%%%%%%%%%%%%%%%%%%%%%%%%%%%%%%%%%%%%%%%%%%%%%%%% 
\vskip -5pt

The resonant peaks described above are illustrated in 
Figure~\ref{fig:InvMass}, where we have set $M(\widetilde{c}_L) = 600$~GeV while 
$B(\widetilde{c}_L \to \mu^+ b) = 1$ as expected in a leptoquark
scenario. The panel on the left corresponds to the 
signal with a muon pair and exactly two jets: the distribution in the 
invariant mass of one of the $\mu b$ pairs corresponding to $0.5 \leq R 
\leq 1.8$ has been shown (empty histogram). The SM background is the 
dark-shaded histogram and the $5\sigma$ fluctuation in the SM background 
is indicated by the light-shaded histogram. Clearly, a huge 
peak -- though not at all a sharp one -- does stand out from the background, indicating a 
squark mass very roughly around 500 GeV. This is about 100 GeV less than the choice 
$M(\widetilde{c}_L) = 600$~GeV, and the discrepancy is clearly due to inaccuracies in jet 
identification and reconstruction. The situation is even better for the central panel, 
which shows the signal with a muon pair and up to three jets, with the same conventions as 
before. In this case, the peak again lies in the 500 -- 550 GeV range, but is now 
significantly higher than the fluctuation in the SM background when
compared with the plot on its left. On the other hand, the cases 
with $b$-tagging, which are virtually background-free and 
are shown in the panel on the right in Figure~\ref{fig:InvMass}, obviously fare best so 
far as significance is concerned. Unfortunately, the signal with a muon pair and exactly two 
$b$ jets is barely discernible, even with 10~fb$^{-1}$ of data. Note that we have set the 
lower edge of the graph at 5 events in 10~fb$^{-1}$ of data. The signal where 
we allow an extra jet in addition to two $b$ jets, however, does much 
better. Obviously, with the huge excesses shown in the first two panels of 
Figure~\ref{fig:InvMass}, discovery of a leptoquark/squark with a 100\% 
branching ratio to $\mu b$ final states is pretty much assured. 
Even with 1~fb$^{-1}$ of data, the significance of the signal can be 
estimated as around $5\sigma$ in the $2\mu$+2~jets
channel, and $6.3\sigma$ in the $2\mu$+$(0 - 3)$~jets channel. 
Marginal improvements in the signal do arise if we relax the 
requirement that $0.5 < R < 1.8$, but eventually this does not change 
the significance by more than 10\%. 

We now come to the case when the $\widetilde{c}_L$ is a squark in 
a RPV model of SUSY with a more natural spectrum than that assumed 
above. As mentioned earlier, in this case, the signal is modified
by ($a$) reduction of the branching ratio $B(\widetilde{c}_L \to \mu b)$
due to the presence of competing channels (see Figure~\ref{fig:clbr})
and ($b$) the production of $\mu$'s and $b$'s through cascade decays of 
other sparticles. Now, two cases arise. If the signal is considered purely 
as an {\it excess} of muons and $b$-jets, the background to this will come 
entirely from the SM, all SUSY contributions being considered part of
the signal. On the other hand, if the presence of {\it resonances} in $\mu b$
final states is considered as the signal (as described above), then the 
production of other SUSY particles through RPC modes will also constitute 
part of the background. Analysis of the first case is described in the next 
section. Here we concentrate on the second option.  

Let us first consider the SUSY backgrounds to the resonant $\widetilde{c}_L$
signals, viz. ($i$) $2\mu + 2b$~jets, ($ii$) $2\mu + 2b$-jets + 0--1 jet,
($iii$) $2\mu + 2$~jets, and ($iv$) $2\mu + 0 - 3$ jets. In our RPV scenario,
SUSY backgrounds will come from the variety of processes listed in Eqn.~(16). 
These are listed for the $\widetilde{c}_L$
squark, but there will be similar processes for the other squarks as well
(except for the first line, of course). A glance at the right-most column of 
Eqn.~(16) will show that all these final states are hadronically rich,
generally having at least four jets. If gluinos are heavier than squarks, 
they will decay to the same squark pairs with the addition of more jets
in the final state. Thus, the signals for the resonant $\mu b$ states 
will come only when there is substantial jet merging. Once subjected 
to the same kinematic cuts and the choice of a restricted band in the 
ratio $R$, as we have done for the signal and SM background, our numerical
analysis shows that the sum total of all these backgrounds contributes only
a minuscule amount to the SM background. Typically, in Figure~\ref{fig:InvMass}
one would add not more than 1 -- 2 events in each bin as shown. We have verified
that this is true for the entire SUSY parameter space of interest. Thus,
the direct effects of the remaining part of the SUSY spectrum are not of much
significance in a search for $\widetilde{c}_L$. Rather, the most important 
effect of having a lighter SUSY spectrum within the 
RPV paradigm will be to reduce the branching ratio 
$B(\widetilde{c}_L \to \mu^+ b)$ from unity due to the opening up of new 
channels such as $\widetilde{c}_L \to c + \widetilde{\chi}_1^0$,  
$\widetilde{c}_L \to s + \widetilde{\chi}_1^+$ and $\widetilde{c}_L \to c + 
\widetilde{g}$, as illustrated in Figure~\ref{fig:clbr}. Since the
four kinds of resonant signal require the presence of  
two $\widetilde{c}_L$'s in the signal, the signal will be suppressed 
by $[B(\widetilde{c}_L \to \mu^+ b)]^2$. A naive appraisal of total
cross-sections (see Figure~\ref{fig:CrossSection}) using only this criterion 
will show that when $B(\widetilde{c}_L \to \mu^+ b)$ falls below about 0.5, it will be difficult to isolate signals above $M(\widetilde{c}_L) \approx 
600 - 700$~GeV.

%%%%%%%%%%%%%%%%%%%%%%%%%%%%%%%%%%%%%%%%%%%%%%%%%%%%%%%%%%%%%%%%%%%%%%%%%%%%%
\begin{figure}[htb]
\centerline{ \epsfxsize= 3.0 in \epsfysize= 3.0 in \epsfbox{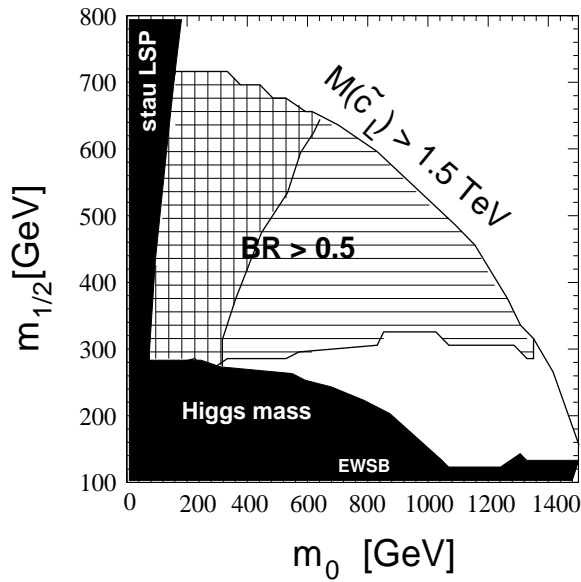} }
\vskip -1pt
\def\baselinestretch{0.8}
\caption{{\footnotesize Illustrating the cMSSM parameter space and the 
region of interest [$M(\widetilde{c}_L) < 1.5$~TeV] for 
squark resonances as discussed in this section. The other cMSSM 
parameters are fixed at $A_0 = 0$, $\tan\beta = 15$ and $\mu > 0$. The 
dark patches are ruled out by theoretical considerations or by direct 
searches as indicated on the Figure. The region hatched with horizontal 
lines has $B(\widetilde{c}_L \to \mu^+ b) > 0.5$ and the region hatched 
with vertical lines has $M(\widetilde{c}_L) < M(\widetilde{g})$. Towards 
the lower right, we have a region with $B(\widetilde{c}_L \to \mu^+ b) < 
0.5$, where the gluino $\widetilde{g}$ is lighter than the squark 
$\widetilde{c}_L$ and hence the partial width for $\widetilde{c}_L \to c + 
\widetilde{g}$ dominates.}}
\def\baselinestretch{1.2}
\label{fig:SUSYMap}
\end{figure}
%%%%%%%%%%%%%%%%%%%%%%%%%%%%%%%%%%%%%%%%%%%%%%%%%%%%%%%%%%%%%%%%%%%%%%%%%%%%%
\vskip -5pt

The situation is similar for studies of the $\mu b$ and $\mu$ + jet 
invariant masses. Obviously the best channels to look for a 
$\widetilde{c}_L$ resonance in the given RPV model seem to be ($i$) the 
total cross-section for muon pairs with up to three jets, and ($ii$) the 
$\mu b$ invariant mass for muon pairs with two $b$-tagged jets and 
possibly a third, untagged jet. In either case, with an integrated
luminosity of 10~fb${-1}$ Figure~\ref{fig:InvMass} shows that a branching ratio 
of $\widetilde{c}_L \to \mu^+ b$ as low as 0.5 can lead to an observable 
signal. The question, therefore, arises: how likely is this branching 
ratio to lie above 0.5? Obviously, the branching ratio will 
depend crucially on the SUSY mass spectrum and the couplings, including 
the mixing angle of gauginos and sfermions. Given the arbitrariness of the 
MSSM spectrum (which is exacerbated in the RPV case) once cannot give a 
definitive answer to this question. However, this issue is partially addressed 
by the plot shown in Figure~\ref{fig:SUSYMap} where we 
assume a minimal supergravity scenario
with radiative electroweak symmetry breaking to get the well-known 
`constrained MSSM' or cMSSM  \cite{mSUGRA, Allanach}. In this 
scenario, we set the trilinear coupling $A_0 = 0$, $\tan\beta = 15$ and 
take the sign of the Higgsino mixing parameter $\mu$ to be positive. 
With this choice of soft SUSY-breaking parameters, the mass spectrum and 
running couplings are calculated at the electroweak scale, using the 
renormalisation group (RG) equations incorporated in the software 
{\sc SuSpect}. We 
then map the region in the $m_0$--$m_{1/2}$ plane which leads to the 
desired branching ratio and exhibit our results in 
Figure~\ref{fig:SUSYMap}. Apart from a small region at the bottom left 
of the panel, the rest of the parameter space is allowed by theoretical 
constraints and data from direct searches at the CERN LEP and Fermilab 
Tevatron. The region indicated with horizontal hatching corresponds, as 
marked, to a branching ratio $B(\widetilde{c}_L \to \mu^+ b) > 0.5$, 
where the signal in Figure~\ref{fig:InvMass} will be discernible at the 
$5\sigma$ level. The overlapping region delineated with vertical 
hatching corresponds to the region where the gluino $\widetilde{g}$ is 
heavier than the $\widetilde{c}_L$, i.e. where gluino cascade decays 
will have a substantial $\widetilde{c}_L$ component. In the unshaded 
region on the lower right side of the graph, it is the $\widetilde{c}_L$ 
which is heavier than the gluino. Here, the dominant decay mode will be 
$\widetilde{c}_L \to c + \widetilde{g}$, as shown in Figure~\ref{fig:clbr}. 
The region outside the rough semicircle marked `$M(\widetilde{c}_L) > 1.5$~TeV' 
is more-or-less 
kinematically disfavoured at the LHC, but its exclusion in the present
graph is due to the fact that the coupling $\lambda'_{223}$ becomes
nonperturbative\footnote{See Eqn.~(\ref{eqn:scaling}) and the discussion 
following it.} in this region.  

It is not difficult to see why the hatched regions have the properties 
indicated above. In general, the gluino mass increases when $m_{1/2}$ 
increases and the squark mass increases when either $m_{1/2}$ or $m_0$ 
(or both) increases. It must be recalled that we assume that the RPV 
coupling $\lambda'_{223}$ increases almost linearly as the mass of 
$\widetilde{c}_L$ increases. If the gluino is heavier than the squark, 
then the squark decays either into the direct RPV channel, which is a 
two-body decay to light states, or through RPC channels into charginos 
and/or neutralinos. If the latter prove to be heavier than the squark, 
they will undergo three-body decays which are suppressed; if they are 
lighter, the available two-body decays will have limited phase space 
when the squark-chargino/neutralino splitting is small. In such cases, 
which are more easily obtained at the left lower side of the Figure, the 
RPV channel dominates i.e. we have the large branching ratio indicated. 
However, as soon as the gluino becomes lighter than the squark, the 
situation changes, since the channel competing with $\widetilde{c}_L 
\toR \mu + b$ is the strong interaction process $\widetilde{c}_L \to c + 
\widetilde{g}$, which becomes the dominant one. This behaviour is 
reproduced qualitatively as we change the other parameters, e.g. 
$\tan\beta$ running up to values as high as 40, $A_0$ of the order of a 
few hundred GeV to a TeV and $\mu <0$. The plot shown here, therefore, 
is, in a sense, the most conservative one.

Finally we ask the question whether one could reproduce the signals 
for $\mu b$ resonances in a model with RPC supersymmetry? Of course,
there are better signals for SUSY in this case \cite{SUSYsearch},
but the question is of academic interest. We note that $pp$ collisions can produce 
a pair of squarks or gluinos, each of which decay to charginos. If a 
muon appears among the decay products of each chargino, there will be 
enough jets in the cascade products of which some may easily be 
$b$-jets. For example, a typical decay chain could be $ \widetilde{q} 
\to q' + \widetilde{\chi}_1^+ \to q' + (\mu^+ \nu_\mu 
\widetilde{\chi}_1^0) \longrightarrow \mu^+ + {\rm jet(s)} + 
\not{\!\!E}_T$, which, repeated on either side with some cancellation of 
missing transverse momenta, could easily mimic the signal we are 
studying. This decay chain could also arise from a gluino: 
$\widetilde{g} \to q + \widetilde{q} \to qq' + \widetilde{\chi}_1^+ \to 
\dots$. Generally RPC supersymmetry signals have a large MET component, 
but this is not true of all the events, and thus we have a residual 
background even after imposing a cut of $\not{\!\!E}_T < 150$~GeV.
Keping this in mind we have carried out a numerical simulation of an
RPC model with a choice of several benchmark points in the parameter
space accessible to LHC \cite{SUSYsearch}. In each case, we find that 
after imposition of all the kinematic cuts and the selection criterion
on the ratio $R$, very few events survive. In the left and middle panels 
of Figure~\ref{fig:InvMass} the deviation from the SM histograms is not
enough to provide even a $2\sigma$ signal, while there are hardly any
events at all in the right-most panel.  

To sum up, therefore, in the RPV model in question, there seems to be a 
strong case for predicting an 
observable squark resonance in the $\mu b$ invariant mass when we look 
at final states with muon pairs accompanied by jets. If the parameters 
are right this can show up simply as an excess in the cross-section for 
a pair of resonances with equal mass. The signal is rather strong in a 
large part of the cMSSM parameter space, fading out somewhat as the 
squark and gluino masses grow heavier, but remaining at some noticeable, 
if not overwhelming, level almost to the kinematic reach of the machine. 
However, if a weak signal, say at the $2\sigma$ level, is seen, this may 
not be a convincing proof of the existence of SUSY with RPV. It would 
require corroboration from other signals. On such possible confirming 
signals we now focus our discussion.

%%%%%%%%%%%%%%%%%%%%%%%%%%%%%%%%%%%%%%%%%%%%%%%%%%%%%%%%%%%%%%%%%%%%%%%%%%%%%%  
\section{The Non-resonant Case}

In case the squark $\widetilde{c}_L$ is produced, but decays mostly 
through RPC channels --- which is what happens when the 
$\widetilde{c}_L$ is heavier than the $\widetilde{g}$ and we have a 
small value of $\lambda'_{223}$ --- RPV will become apparent only at the 
last stage in the SUSY cascade decay, i.e. in the decay of the lightest 
supersymmetric particle (LSP) into a muon plus two jets. 
In Eq.~(16), this will correspond to the 
decay modes indicated between parentheses in the extreme right column. 
As we have already noted, such final states arise at least 70\% 
of the time and always contain a muon of indeterminate sign, together with 
multiple jets. When a gluon or a 
squark other than the $\widetilde{c}_L$ is produced, the decay channels 
will be very similar, except for the direct decay into a $\mu b$ pair. 
These will decay through RPC cascades into the LSP 
$\widetilde{\chi}_1^0$, which will then decay into a muon, a $b$ jet as 
well as other jets and MET. When a pair of gluinos or 
squarks is produced, these will be produced with double the multiplicity 
for every particle, including jets. Hence, even if we do not have an 
observable $\mu b$ resonance, an excess of $\mu$'s and $b$-jets in the 
final state will be a clear indicator of the RPV model in question.

It may be noted that the signals discussed in this section are not so 
crucially dependent on the actual value of the $\lambda'_{223}$ 
coupling, since the LSP has essentially only one decay mode. This 
discussion will, therefore, be largely valid 
{\it even if we do not impose a scaling of 
$\lambda'_{223}$ with $M(\widetilde{c}_L)$} as was done in the previous 
section. This is an example of an analysis which holds irrespective of 
whether we correlate low-energy effects with the high-energy frontier, 
and is, in a sense, more robust than the signals discussed in the 
previous section.

To showcase the above, we choose a particular signal out of the many 
possible ones. We trigger on a final state with a muon pair 
(irrespective of sign), only one tagged $b$-jet and a complement of {\it 
at least} 3 more jets, making a total of 4 jets. This final state will 
receive contributions from almost all the cascade decays ending in an 
LSP listed in Eq.~(16). The resonant signal, when it dominates, will 
also contribute to this final state, when it arises from the decay of a 
pair of gluinos, but not, in general, if a $\widetilde{c}_L 
{\widetilde{c}_L^{\,\ast}}$ pair is produced directly. Some small 
contributions may still arise in the last case if extra jets are 
generated by radiative or jet-splitting effects, but these will not 
change the qualitative features of this signal, and in any case, are 
taken care of in our analysis which uses the event generator {\sc 
pythia}.

The above signal is not, however, free from backgrounds. The principal 
background will arise from $t\bar{t}$ production, with its huge NLO 
cross-section: almost up to a nanobarn. If both the top quarks decay 
semileptonically, to a muon, a $b$-jet and missing energy, we shall have 
a final state short by two jets from the given signal. Such extra jets 
can easily be generated by radiative or jet-splitting effects, as 
mentioned above, and though it will lead to an automatic suppression by 
at least ${\cal O}(\alpha_s^2)$, as a fraction of the enormous 
$t\bar{t}$ cross-section it will still be a formidable background to the 
signal under consideration. Other obvious contributors will be a final 
state with a $Z$ boson and four jets, where at least one of the jets is 
a $b$ jet and the $Z$ decays to a $\mu^+\mu^-$ pair, or a $W^+W^-$ plus 
four jets, where both $W$'s decay to muons and at least one of the jets 
is a $b$ jet. These are much smaller, however, than the $t\bar{t}$ 
contribution, and will not affect the signal overmuch. It is true that 
in all these effects, the muon-$p_T$ will be somewhat restricted by the 
fact that it arises from the decay of an on-shell $W/Z$ boson, but this 
is not of much help, since the muons in the signal, except for the very 
few which arise from the decay of a resonant $\widetilde{c}_L$, are a 
product of the decay of the LSP, which may not be significantly heavier 
than the $W$ boson. Moreover, the $p_T$ distribution will be much 
smeared-out by the kinematics of the rather complicated final state, and 
may not look very different, whether the source is a $t\bar{t}$ pair, or 
a SUSY cascade decay.

\begin{figure}[htb]
\centerline{ \epsfxsize= 3.0 in \epsfysize= 2.9 in \epsfbox{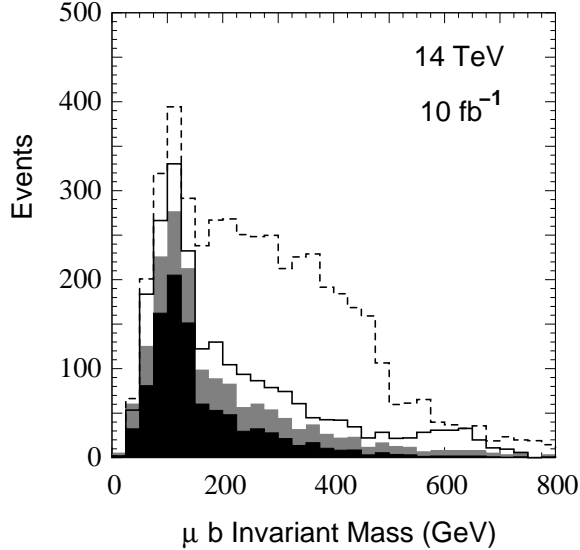} }
\vskip -1pt
\def\baselinestretch{0.8}
\caption{{\footnotesize $\mu b$ invariant mass distribution for the 
signal with a muon pair and $\geq 4$ jets of which at least one is a 
$b$-jet. The solid dark-shaded histogram represents the SM background, 
while the solid light-shaded histogram represents its possible 
fluctuations at the $5\sigma$ level. The unshaded histogram with a solid 
line represents the RPV signal when the gluino is heavier than the 
$\widetilde{c}_L$, while the unshaded histogram with the broken line 
represents the RPV signal when the $\widetilde{c}_L$ is heavier than the 
gluino. Values of the SUSY parameters in both cases are given in the 
text.}}
\def\baselinestretch{1.2}
\label{fig:InvMass4}
\end{figure}

The viability of this signal is illustrated in Figure~\ref{fig:InvMass4}.
In our numerical analysis, the following kinematic cuts were imposed.
\vspace*{-0.2in}
\begin{enumerate}

\item The muons were required to have transverse momentum $p_T(\mu) > 
20$~GeV.

\item The jets were ordered as ${\rm J}_1, {\rm J}_2, \dots$ according 
to their $p_T$ values, after which we imposed
$$
p_T({\rm J}_1) > 100~{\rm GeV} 
\qquad\qquad
p_T({\rm J}_i) > 50~{\rm GeV~for}~i = 2,3,4,\dots 
$$

\item The transverse sphericity $\sigma$ of each event was required to 
satisfy $\sigma > 0.2$.
\end{enumerate}
\vspace*{-0.2in}
No restrictions were made on the missing energy and momentum of the 
events; likewise, the pairwise invariant masses were allowed to vary 
freely. Finally, we plot the $\mu b$ invariant mass, taking the harder 
of the two muons in the signal\footnote{We have checked that taking the 
softer muon does not change the qualitative features of the graph.}.

%%%%%%%%%%%%%%%%%%%%%%%%%%%%%%%%%%%%%%%%%%%%%%%%%%%%%%%%%%%%%%%%%%%%%%%%%%%%%
%%%%%%%%%%%%%%%%%%%%%%%%%%%%%%%%%%%%%%%%%%%%%%%%%%%%%%%%%%%%%%%%%%%%%%%%%%%%%
\vskip -5pt

In Figure~\ref{fig:InvMass4}, as in Figure~\ref{fig:InvMass}, the 
dark-shaded histogram represents the SM background, principally 
originating from $t\bar{t}$ production, and the light-shaded histogram 
represents its possible $5\sigma$ fluctuations. The unshaded histogram 
drawn with a solid line represents the prediction of the RPV model, 
where the entire SUSY spectrum is generated in a cMSSM with $m_0 
= 100$~GeV, $m_{1/2} = 300$~GeV, $A_0 = 0$, $\tan\beta = 15$ and 
sgn($\mu$) = +1. In this scenario, we predict $M(\widetilde{c}_L) \simeq 
650$~GeV and $M(\widetilde{g}) \simeq 740$~GeV. To keep the low-energy 
solutions consistent, we scale $\lambda'_{223} = 0.65$, which is 
comparable to the electroweak SU(2)$_{\rm L}$ coupling $g$. This set of 
parameters determines everything in the model, from production 
cross-sections to branching ratios, and the above plot shows that 
we predict 
a clear signal above the $5\sigma$ discovery level at the LHC. The 
signal shows a peaking behaviour around 100~GeV, which may be attributed 
to the cases where the muon and $b$-jet arise from the decay of the 
neutralino $\widetilde{\chi}_1^0$ LSP, which has a mass in just that 
ballpark. There is also a modest peaking around 600~GeV, which may be 
attributed to the $\widetilde{c}_L$ resonance, and is the analogue, for 
this more-inclusive signal, of the sharper peaks shown in 
Figure~\ref{fig:InvMass}. The number of events in this region, with the 
luminosity projection of 10~fb$^{-1}$ is much larger than the 
background, and hence, if seen, would be a `smoking gun' signal for new 
physics beyond the SM.

If we consider a situation where the gluino is lighter than the squark 
$\widetilde{c}_L$ we should expect no such peak around 
$M(\widetilde{c}_L)$, and indeed, this is the case with the unshaded 
histogram drawn with broken lines in Figure~\ref{fig:InvMass4}. In this 
case, all the parameters have been kept at the same values as before, 
with the exception of the gluino mass, which has been reduced (by hand, 
as it were) from 740~GeV to 620~GeV\footnote{Obviously, this is not 
consistent with the cMSSM spectrum, but it can be generated easily 
enough in a non-universal scenario where the SU(3)$_{\rm c}$ gaugino 
mass is taken different from those in the electroweak sector.}. At 
620~GeV, the gluino is lighter than the $\widetilde{c}_L$ at 650~GeV, 
and hence the $\widetilde{c}_L$ will decay principally into $c + 
\widetilde{g}$. The cascade decays will now resemble the RPC scenario, 
with an enhanced gluino production due to the smaller mass, and all 
these decays will end in $\widetilde{\chi}_1^0$ LSPs, which will then 
decay through $\lambda'_{223}$. The muon and $b$-jet will thus arise 
from these end-products of SUSY cascades, which explains why we get 
copious numbers of signal events as shown by the histogram in 
Figure~\ref{fig:InvMass4}. Of course, the huge increase in the signal 
may be attributed in increased pair-production of the lighter gluino in 
this case, and may not be as large over the full cMSSM parameter 
space, but the point which we would prefer to stress is that we get a 
substantial signal in {\it both} scenarios, whether there is a 
$\widetilde{c}_L$ resonance or not. This signal will fade out as the 
squarks and gluinos grow heavier, as is true with all SUSY signals at 
the LHC, but in this case, it is not sensible to go to very high values 
of the $\widetilde{c}_L$ mass, since then one cannot get large FCNC 
effects without pushing $\lambda'_{223}$ towards the non-perturbative 
regime. For $M(\widetilde{c}_L) \, \gsim \, 1$~TeV, therefore, it is 
doubtful if RPV can simultaneously be 
observed in low-energy processes and at the LHC.

It is interesting to ask how we can tell if the non-resonant signal 
considered in this section arises from SUSY of the RPC or RPV type. 
Actually this distinction is very easy to make, for the RPC model will
produce roughly equal numbers of $e^\pm$ and $\mu^\pm$ in the final 
state, since these will mainly arise as decay products of $W,Z$ bosons
or near-degenerate sleptons.
On the other hand, RPV decays of the $\widetilde{\chi}_1^0$  LSP 
through a $\lambda^\prime_{223}$ coupling
will inevitably produce a huge excess of muon final states as compared
with electron final states. This is illustrated in Table~1 below, where
the number of events 
is exhibited with both RPC and RPV cases, for $ee$, $e\mu$ and $\mu\mu$
(of arbitrary sign) final states accompanied by jets and MET, as considered
above. We have also imposed the additional cut $\not{\!\!E}_T < 150$~GeV,
which helps to suppress the RPC signal compared with the RPV one.

\vspace*{-0.1in}
\begin{table}[h]
\begin{center}
\begin{tabular}{rccc}
     & $N_{ee}$ & $N_{e\mu}/N_{ee}$ & $N_{\mu\mu}/N_{ee}$  \\ \hline\hline
RPC  &  30      &  1.1              & 1.15                  \\ 
RPV  &  53      &  7.5              & 41.0                 \\ 
\hline
\end{tabular}
\def\baselinestretch{0.8}
\caption{{\footnotesize Comparing similar final states with electrons and muons in the case of RPC and RPV varieties of SUSY. The numbers $N$
correspond only to the signal. Parameter choices are
the same as used to generate the solid line histogram in Figure~\ref{fig:InvMass4}. We take $\sqrt{s} = 14$~TeV and ${\cal L} = 10$~fb$^{-1}$
for our simulation. }}
\def\baselinestretch{1.2}
\label{tab:rpcrpv}
\end{center}
\end{table}
\vspace*{-0.25in}
It is obvious from Table~\ref{tab:rpcrpv} that irrespective of the actual
numbers, the flavour ratios obey the hierarchy
\begin{equation}
\left(\frac{N_{e\mu}}{N_{ee}}\right)_{\rm RPV} 
> \left(\frac{N_{e\mu}}{N_{ee}}\right)_{\rm RPC}
\qquad\qquad\qquad 
\left(\frac{N_{\mu\mu}}{N_{ee}}\right)_{\rm RPV} 
\gg \left(\frac{N_{\mu\mu}}{N_{ee}}\right)_{\rm RPC}
\label{eqn:muonratio}
\end{equation}
i.e. there is a huge relative excess 
in muonic final states compared to electronic final states in the RPV case.
While the ratios exhibited in Table~1 will vary considerably when we 
move to other points in the parameter space, the {\it qualitative} feature
in Eqn.~(\ref{eqn:muonratio})
will generically be reproduced, and may be taken as a `smoking gun'
signal for R-parity violation through a $\lambda^\prime_{223}$ coupling.

%%%%%%%%%%%%%%%%%%%%%%%%%%%%%%%%%%%%%%%%%%%%%%%%%%%%%%%%%%%%%%%%%%%%%%%%%%%%%%
\section{Same-Sign Leptons and their Significance}

In the two previous sections we have discussed searches for RPV in 
certain final states and invariant mass distributions, using mainly the 
kinematic properties of the SUSY particle spectrum. We now turn to an 
important {\it dynamic} property, viz. the fact that the charge-neutral 
gauginos -- gluinos and neutralinos -- are fermions of Majorana type, 
i.e. they are their own antiparticles. No such particles are known to 
exist in the SM; in fact, a multiplicity of Majorana particles is a 
special feature of SUSY models. In the context of our analysis, 
therefore, we should expect some unique features arising from the 
Majorana property and it turns out that we can expect some really 
spectacular signals which are the subject of discussion in the current 
section.

%%%%%%%%%%%%%%%%%%%%%%%%%%%%%%%%%%%%%%%%%%%%%%%%%%%%%%%%%%%%%%%%%%%%%%%%%%%%%
\begin{figure}[htb]
\centerline{ \epsfxsize= 4.5 in \epsfysize= 5.2 in \epsfbox{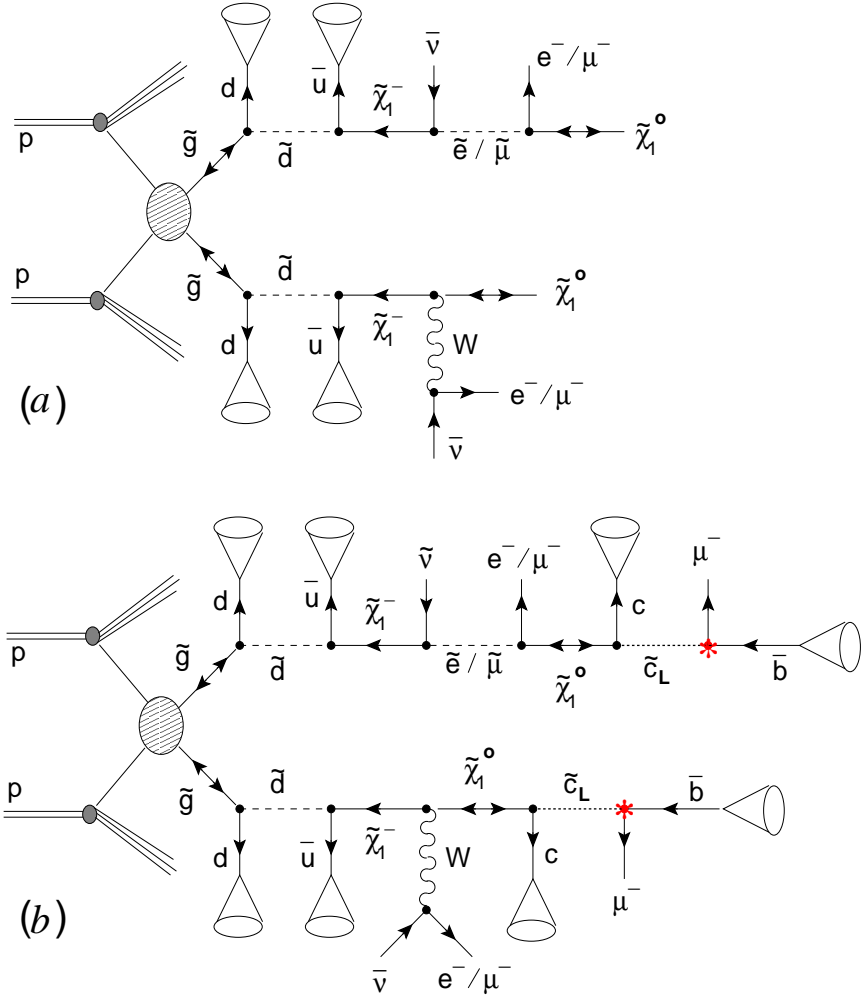} }
\vskip -1pt
\def\baselinestretch{0.8}
\caption{{\footnotesize Same-sign lepton production at the LHC in ($a$) 
RPC and ($b$) RPV models. In ($a$) the final state is a same-sign lepton 
pair (SSLP), together with jets and MET; in ($b$) the final state is a 
same-sign lepton quartet (SSLQ), together with jets and MET. The RPV 
couplings correspond to the vertices marked with asterisks on the 
extreme right in ($b$). Little cones indicate jets and the dotted (not 
dashed) lines indicate virtual (off-shell) states. In ($b$), if the $W$ 
boson decays hadronically, the final state will be a same-sign lepton 
triad (SSLT), together with jets and MET.
}}
\def\baselinestretch{1.2}
\label{fig:feynssl}
\end{figure}
%%%%%%%%%%%%%%%%%%%%%%%%%%%%%%%%%%%%%%%%%%%%%%%%%%%%%%%%%%%%%%%%%%%%%%%%%%%%%
\vskip -5pt

Before discussing the specific effects in the current RPV model, let us 
briefly review the predictions of the RPC version of SUSY with regard to 
Majorana fermions. This is illustrated in Figure~\ref{fig:feynssl}($a$) 
as arising from the gluino decay chain
$$
\widetilde{g} \longrightarrow q + \widetilde{q} 
\longrightarrow q + q' + \widetilde{\chi}_1^\pm 
\longrightarrow q + q' + \ell^\pm + \not{\!\!E}_T
$$  
where the decay of the chargino can take place through two different 
channels as shown. In the Figure, the little cones indicate hadronic 
jets and the clashing arrows indicate Majorana fermions. The Majorana 
nature of the neutralino LSP $\widetilde{\chi}_1^0$, however, has no 
effect, since it contributes only to MET. However, the gluinos 
$\widetilde{g}$ can give rise to identical charginos on either side, 
leading to same-sign lepton pairs (SSLP). Here, as before, we consider 
only the charged leptons $e^\pm$ and $\mu^\pm$. Of course, there can 
also be opposite-sign leptons in these decay chains, but these will have 
large SM backgrounds, whereas there is hardly any SM background to the 
same sign dilepton signal. Note that unless there is a large mass-splitting
between the selectron and the smuon, the probability for a chargino to decay to
an electron is roughly the same as its probability to decay to a muon,
and hence same-sign $ee$, $\mu\mu$ and $e\mu$ 
pairs would arise in the rough proportions 1:1:2 in the RPC version of
SUSY. If we consider only gluino production, then $\ell^+\ell^+$ and
$\ell^-\ell^-$ pairs will be produced in equal proportions, but 
when squark pair production or squark-gluino pairs are 
considered at a $pp$ collider, $\ell^+\ell^+$ pairs dominate over 
$\ell^-\ell^-$ pairs \cite{cascades}. The flavour ratio is, however, maintained,
irrespective of sign, and this is what our analysis will focus upon.
 
The situation becomes much more interesting if R parity is violated \cite{HOD}. 
For, as we have seen in this case, the neutralino LSP 
$\widetilde{\chi}_1^0$ can decay through a virtual $\widetilde{c}_L$ 
into a final state with a $\mu^\pm$ and two jets. Both signs of the muon 
are permitted, because of the Majorana nature of the 
$\widetilde{\chi}_1^0$, which here assumes an important role, unlike the 
previous case.  The decay chain shown in Figure~\ref{fig:feynssl}($a$) 
is now extended, as shown in Figure~\ref{fig:feynssl}($b$), by the decay 
of the LSP's, and it can be seen, that, in addition to the $e^\pm 
e^\pm$, $\mu^\pm\mu^\pm$ and $e^\pm \mu^\pm$ pairs arising from the 
chargino decay, we can add on a same-sign $\mu^\pm\mu^\pm$ pair from the 
neutralino decay --- the result being same-sign lepton {\it quartets} 
(SSLQ) such as $e^-e^-\mu^-\mu^-$ or $e^-\mu^-\mu^-\mu^-$ or even 
$\mu^-\mu^-\mu^-\mu^-$ (and, of course, their positively-charged 
counterparts). It is also possible for one or more of the charginos to 
decay hadronically, i.e. $$ \widetilde{\chi}_1^\pm \to 
\widetilde{\chi}_1^0 + W^\pm \to \widetilde{\chi}_1^0 + q\bar{q} $$ 
Insertion of this in place of {\it one} of the chargino decays shown in 
Figure~\ref{fig:feynssl}($b$) would lead to a prediction of same-sign 
lepton {\it triads}, (SSLT) of the form $e^-e^-\mu^-$, $e^-\mu^-\mu^-$ 
or $\mu^-\mu^-\mu^-$ (and their positively-charged counterparts). In 
fact, since the hadronic branching ratio of the chargino is generally 
larger than its branching ratio to leptons $e, \mu$, one may expect many 
more events with SSLT than SSLQ. The most probable channel will be for 
both the charginos to decay hadronically, in which case, it is still 
possible to have SSLP --- $\mu^\pm\mu^\pm$ --- from the decays of the 
neutralinos. Thus, in an RPV model one would predict \cite{HOD} 
the simultaneous existence of SSLP, SSLT and SSLQ, in decreasing order of 
cross-section.  
All have vanishingly small backgrounds in the SM --- typically below
$10^{-1}$~fb for SSLT and even lower for SSLQ, as shown in 
Ref.~\cite{HOD}. In the RPC versions of SUSY, all of SSLP, SSLT and
SSLQ will arise, but, as discussed above, the 
number of muonic events will be roughly equal to the number of 
electronic events for the RPC case, unlike the RPV case considered
here, where muons will be overwhelmingly dominant. We shall take
up this point in more detail in the following discussion. 

We have made a numerical study of the SSLP, SSLT and SSLQ signals at the 
LHC, using, as before, the event generator {\sc pythia}. For these 
studies at $\sqrt{s} = 14$~TeV, the following kinematic cuts were 
imposed:
\vspace*{-0.2in}
\begin{enumerate}
\item For the SSLP case, we imposed 
$$
p_T(\ell) > 20~{\rm GeV}
$$
for both the leptons. Up to 4 accompanying jets were considered, and the 
jets were ordered as ${\rm J}_1, {\rm J}_2, \dots$ according to their 
$p_T$ values, after which we imposed $$ p_T^{{\rm J}_1} > 100~{\rm GeV} 
\qquad\qquad p_T^{{\rm J}_i} > 50~{\rm GeV~for}~i = 2,3,4. $$ The 
transverse sphericity $\sigma$ of each event was also required to 
satisfy $\sigma > 0.2$. No cut was imposed on the MET.

\item For the SSLT case, we imposed a cut
$$
p_T(\ell) > 20~{\rm GeV}
$$
for all three leptons, and no other cuts on jets and/or MET. 

\item For the SSLQ case, the cut on the lepton transverse momentum was 
reduced to 
$$
p_T^\ell > 10~{\rm GeV} \ .
$$
\end{enumerate}
\vspace*{-0.2in}
For studies at 7~TeV, we imposed $p_T^\ell > 10~{\rm GeV}$ for both SSLT 
and SSLQ signals.

%%%%%%%%%%%%%%%%%%%%%%%%%%%%%%%%%%%%%%%%%%%%%%%%%%%%%%%%%%%%%%%%%%%%%%%%%%%%%
\begin{figure}[htb]
\centerline{ \epsfxsize= 5.2 in \epsfysize= 2.5 in \epsfbox{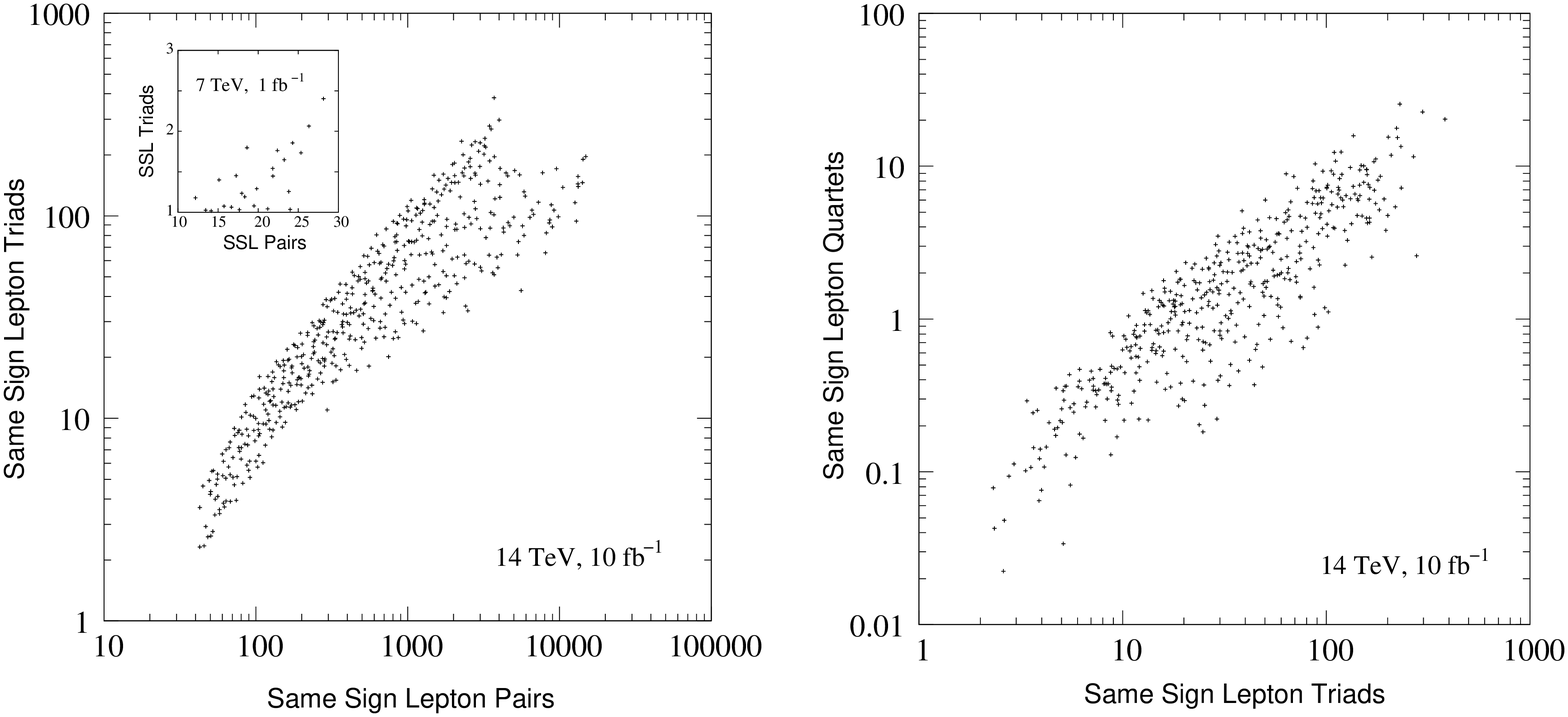} }
\vskip -1pt
\def\baselinestretch{0.8}
\caption{{\footnotesize Same sign lepton production at the LHC for 
10~fb$^{-1}$ luminosity at the 14~TeV run. The left panel shows the 
number of same-sign triads (SSL) versus same sign pairs (SSLP), while the 
right panel shows the number of same-sign quartets (SSLQ) versus triads 
(SSLT). Each point represents a set of vales of the cMSSM parameter 
space. The inset box shows the SSLT-SSLP plot for 1~fb$^{-1}$ luminosity 
at the 7~TeV run.
}}
\def\baselinestretch{1.2}
\label{fig:ssl}
\end{figure}
%%%%%%%%%%%%%%%%%%%%%%%%%%%%%%%%%%%%%%%%%%%%%%%%%%%%%%%%%%%%%%%%%%%%%%%%%%%%%
\vskip -5pt

Our results are presented in Figure~\ref{fig:ssl}. In the left panel, we 
plot, for $\sqrt{s} = 14$~TeV and an integrated luminosity of 
10~fb$^{-1}$, the number of events with SSLT versus the number of events 
with SSLP, while the parameters of the model are sampled randomly over 
the ranges:
$$
m_0 = 0.1 - 2.0~{\rm TeV}
\qquad\qquad
m_{1/2} = 0.1 - 1.5~{\rm TeV}
\qquad\qquad
A_0 = 0, \ \tan\beta = 15, \ \mu > 0 \ .
$$
The mass spectrum for this model is 
generated as before, assuming the cMSSM. Given this model, each point on 
this plane represents a particular choice of $m_0$, $m_{1/2}$ and 
$\tan\beta$, while we set $A_0 = 0$ and take $\mu > 0$. The corresponding
mass spectrum is required to satisfy LEP constraints on the light Higgs
boson mass as well as the requirement that $M(\widetilde{\chi}_L) < 1.5$~TeV,
which defines our region of interest. It may be seen 
that the predicted points lie in the neighbourhood of a straight line with 
slope of approximately 0.1 --- this is just the ratio between the leptonic and 
hadronic branching ratios of the chargino, as the above discussion would 
lead us to expect. What is more important, however, is that fact that 
Figure~\ref{fig:ssl} tells us that for every 10 SSLP events seen, we should 
expect at least one SSLT event. The numbers could be as high, for some 
regions of the parameter space, as a few thousand SSLP events and around 
a hundred SSLT events.

The situation for $\sqrt{s} = 7$~TeV and a data sample of 1~fb$^{-1}$ is 
shown in the box inset on the left side of Figure~\ref{fig:ssl}. The 
sparseness of points is an artifact of the fact that the whole parameter 
space was scanned with the same resolution as in the 14~TeV case, out of which 
just a small sliver contributes to the 7~TeV signal. What is exciting, 
however, is that it is possible, if the parameters happen to be just 
right, that we could see some 10--25 SSLP events accompanied by just one 
or two SSLT events. Should such a signal be seen, we would have found 
solid evidence for SUSY, and just a tantalizing hint for RPV. This is as 
far as the 7~TeV run can go; clinching evidence will have to wait for 
the 14~TeV run.

If the SSLT events are roughly an order of magnitude rarer than the SSLP 
events, then we should expect the SSLQ events to be another order of 
magnitude rarer still. This is, in fact, the case, as is shown in the 
right panel in Figure~\ref{fig:ssl}. SSLQ events are not predicted at 
the 7~TeV run, so this panel is exclusively for the $\sqrt{s} = 14$~TeV 
and 10~fb$^{-1}$ run. Obviously, if we see less than 10 SSLT events, we 
should not expect to see any SSLQ events, but, as we have seen before, 
if the parameter space turns out to be favourable, we may see as many as 
a hundred SSLT events with up to 10 SSLQ events. Such spectacular signals 
would immediately indicate the presence of RPV, as envisaged in this 
model.

We now come to the issue of distinction of SSLP and perhaps SSLT signals
arising in this model versus those arising in conventional RPC SUSY. 
Studies of the latter has been carried out in Ref.~\cite{cascades}
and, more recently, in Ref.~\cite{SSL}, and
it is not our purpose, in this work, to repeat such analyses. Rather, we 
take up the question: if we do indeed observe SSLP and SSLT 
events\footnote{SSLQ events are a rare option that almost never arises in 
RPC SUSY for an integrated luminosity of 10~fb$^{-1}$.} how can we tell
if the underlying SUSY model is of the RPC or RPV type? The answer is 
actually very simple: as we have seen, in RPC versions of SUSY, cascade
decays tend to end in equal numbers of electrons and muons, whereas in
our RPV model with a large $\lambda'_{223}$ coupling, 
muons have a tendency to dominate. This has already shown
up in a different context in Table~1, but in the case of same-sign leptons
it can be used in an equally dramatic way. 

%%%%%%%%%%%%%%%%%%%%%%%%%%%%%%%%%%%%%%%%%%%%%%%%%%%%%%%%%%%%%%%%%%%%%%%%%%%%%
\begin{figure}[htb]
\centerline{ \epsfxsize= 5.2 in \epsfysize= 2.5 in \epsfbox{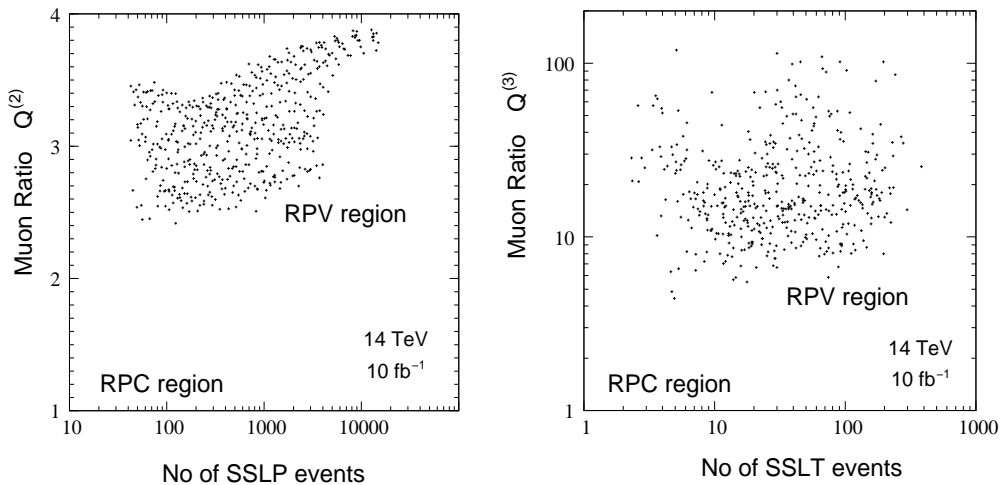} }
\vskip -1pt
\def\baselinestretch{0.8}
\caption{{\footnotesize `Muon Ratios' constructed for the SSLP and SSLT
samples for the RPV model. The parameter ranges are identical with those 
of Figure~\ref{fig:ssl}. The left panel is for SSLP signals, whereas the
right panel is for SSLT signals. Note the clear separation between the RPV 
and RPC regions indicated on both panels. }}
\def\baselinestretch{1.2}
\label{fig:sslratio}
\end{figure}
%%%%%%%%%%%%%%%%%%%%%%%%%%%%%%%%%%%%%%%%%%%%%%%%%%%%%%%%%%%%%%%%%%%%%%%%%%%%%
\vskip -5pt

Taking up the case of SSLP first, we have already noted that same-sign 
$ee$, $\mu\mu$ and $e\mu$ pairs would arise in the rough proportions 1:1:2.
This encourages us to construct the `muon ratio'
\begin{equation}  
Q^{(2)} = \frac{4\sigma^{\rm SS}_{\mu\mu}} 
{\sigma^{\rm SS}_{ee} + \sigma^{\rm SS}_{e\mu} + \sigma^{\rm SS}_{\mu\mu}}
\end{equation}
and predict $Q^{(2)} \approx 1$ in the RPC model. Similarly, for the 
SSLT case, we can construct a `muon ratio'
\begin{equation}  
Q^{(3)} = \frac{\sigma^{\rm SS}_{\mu\mu\mu} + \sigma^{\rm SS}_{e\mu\mu}} 
{\sigma^{\rm SS}_{eee} + \sigma^{\rm SS}_{ee\mu} }
\end{equation}
with an RPC prediction $Q^{(3)} \approx 1$. It only remains, then, to
see what these ratios work out to in the RPV model, and this is exhibited
in Figure~\ref{fig:sslratio}. It may be noted that within the given 
parameter choice, there will always be at least 40 SSLP events, which 
allows a meaningful construction of the $Q^{(2)}$ variable, which is 
seen never to drop below 2.4. The situation is not so clear for SSLT
events, where the prediction drops to as low as 2 events, where, obviously,
construction of the $Q^{(3)}$ variable is not meaningful (unless the 
luminosity goes up by about an order of magnitude). However, for the 
part of the parameter space where we get 10 or more events, the `muon
ratio' $Q^{(3)}$ never drops below 5, which is a far cry from the RPC
prediction -- close to unity. Thus, distinction between RPC SUSY and the 
model discussed in this work is a simple matter for SSLP signals and 
for SSLT signals if the parameters are favourable.    

%%%%%%%%%%%%%%%%%%%%%%%%%%%%%%%%%%%%%%%%%%%%%%%%%%%%%%%%%%%%%%%%%%%%%%%%%%%%%
\begin{figure}[htb]
\centerline{ \epsfxsize= 5.2 in \epsfysize= 2.5 in \epsfbox{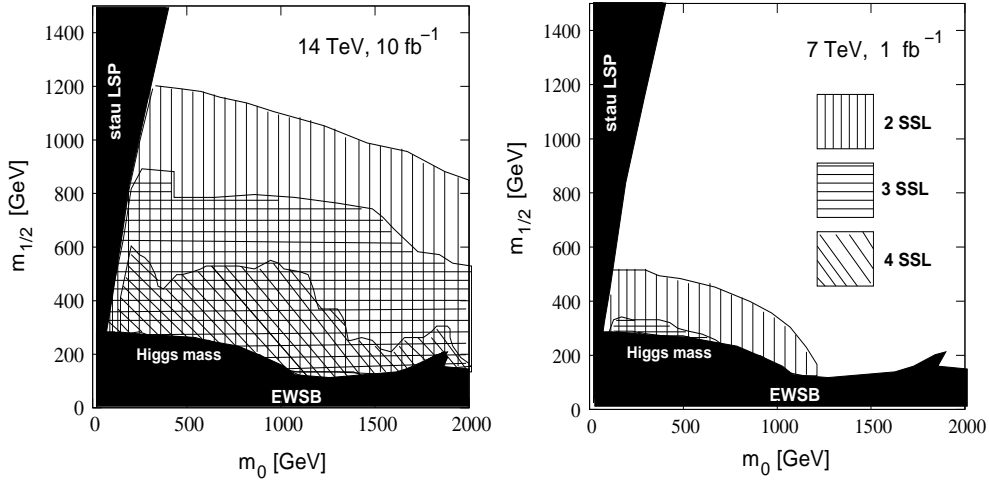} }
\vskip -1pt
\def\baselinestretch{0.8}
\caption{{\footnotesize Same-sign lepton production at the LHC. The left 
panel corresponds to 10~fb$^{-1}$ luminosity at the 14~TeV run and the 
right panel corresponds to 1~fb$^{-1}$ luminosity at the 7~TeV run. Dark 
regions are ruled out by theoretical considerations or by direct 
searches. Regions hatched with vertical, horizontal and slanting lines 
correspond to viable signals with lepton pairs, triads and quartets 
respectively.}}
\def\baselinestretch{1.2}
\label{fig:sslmap}
\end{figure}
%%%%%%%%%%%%%%%%%%%%%%%%%%%%%%%%%%%%%%%%%%%%%%%%%%%%%%%%%%%%%%%%%%%%%%%%%%%%%
\vskip -5pt

The above discussion would be incomplete without a discussion of which 
part of the parameter space in this model would be excluded if 
SSL signals are not seen at the LHC. This is shown in 
Figure~\ref{fig:sslmap}. The left panel corresponds to the $\sqrt{s} = 
14$~TeV and 10~fb$^{-1}$ run, while the right panel is for the $\sqrt{s} 
= 7$~TeV and 1~fb$^{-1}$ run. As before, we work in the cMSSM framework 
and show the $m_0$--$m_{1/2}$ plane with $\tan\beta = 15$, $A_0 = 0$ and 
$\mu > 0$. The dark shaded regions are ruled out by theoretical 
considerations or by direct searches, as marked on the Figure. The same 
conventions are followed for both panels. The regions where one can 
expect at least {\it one} SSLP event are indicated by vertical hatching, 
the regions corresponding to at least {\it one} SSLT event are indicated 
by horizontal hatching and the region (there is no such region in the 
7~TeV case) corresponding to at least {\it one} SSLQ event are indicated 
by slanting hatching.

Even a glance at Figure~\ref{fig:sslmap} will show that SSLP events are 
predicted over most of the kinematically accessible parameter space of 
the cMSSM-based RPV scenario at the LHC. The region thus mapped (vertical 
hatching) is somewhat more extensive than in the RPC case, because that 
requires the charginos to decay to leptons rather than jets. However,
the best exclusion plots in the RPC case are obtained not from SSL signals
but from a 4-jets plus MET signal \cite{SUSYsearch}. In our RPV 
model the origin of SSLP is from the neutralino decay which has no 
competing channel (except the RPV one to $\nu_\mu s\bar{b}$). However, 
mere discovery of SSLP events with a somewhat larger cross-section could 
perhaps be explained by some variant of the cMSSM. What is more 
interesting is that SSLT events are also predicted over a wide swath of 
the parameter space, and these could prove the clinching evidence for 
RPV as a phenomenon. If the parameters are very favourable, i.e. we have 
a SUSY-around-the-corner scenario, we may eventually be able to see SSLQ 
events as well. Such events would surely satisfy even the most die-hard 
sceptic of the existence of RPV. However, as the corresponding patch in 
the parameter space is small, one should not bank upon it. Similarly, 
SSL signals will appear only for a very small patch of the parameter 
space in the panel on the right, i.e. for the 7~TeV run, and this 
indicates that an RPV discovery there would require the parameters to be 
in just the right ballpark.

%%%%%%%%%%%%%%%%%%%%%%%%%%%%%%%%%%%%%%%%%%%%%%%%%%%%%%%%%%%%%%%%%%%%%%%%%%%%%%
\section{Summary and Conclusions}

In this work, we build on the premise that new physics beyond the 
Standard Model, may affect low-energy processes and at the same time 
lead to direct searches at the LHC. While it is true that there exists 
no measurement at low energies which is in direct contradiction with the 
Standard Model prediction within a reasonable level of experimental 
error, it is also true that we are on the threshold of making low-energy 
measurements, which may well be much larger than the Standard Model 
estimates, and would, therefore, be better understood if we assume the 
existence of new physics. We focus on a set of such phenomena, viz., 
neutrino masses, rare branching ratios of the tau lepton, the phase in 
$B_s$--$\bar{B}_s$ mixing, and the decay $D_s \to \ell\nu$, which could 
collectively indicate the existence of a leptoquark with equal 
$\lambda'_{223}$ and $\lambda'_{323}$ couplings, or alternatively, a 
supersymmetric model where R-parity is violated by just this pair of 
couplings, without falling foul of the muon anomalous magnetic moment. 
Current experimental data are compatible with a coupling strength 
$\lambda'/\widetilde{m} \approx 1$~TeV$^{-1}$, where $\widetilde{m}$ is 
the mass of the leptoquark or the R-parity-violating squark. As this is 
just the scale at which the LHC processes are due to take place, we then 
set out to explore the consequences of having such a model in the 
context of LHC signals.

We have explored three kinds of LHC signals. The first would be 
important if the new physics consists of a leptoquark, with only the 
couplings mentioned above. These would be produced copiously at the LHC 
because of their strong interactions, and their characteristic decays to 
a lepton and a jet could then be exploited to construct pairs of 
resonances, whose observation at the LHC would be a simple matter, as we 
demonstrate by a detailed Monte Carlo simulation of the signal 
vis-\'a-vis its principal backgrounds. If, however, the new physics is 
not just a leptoquark, but supersymmetry with R-parity violation, the 
situation becomes much more complicated, --- as is to be expected when 
the particle spectrum is augmented by twenty-eight new 
particles\footnote{thirty-one if there are right-handed neutrinos.} and 
a host of new interactions. On the negative side, the R-parity-violating 
squarks, which resemble leptoquarks, would have competing decay 
channels, weakening the resonance signals, and we demonstrate that there 
are indeed parts of the parameter space where such signals are no longer 
viable.

Supersymmetry, however, provides new detection channels, whether 
R-parity is conserved or not. Keeping in mind the part of the parameter 
space where leptoquark-like resonances would go undetected, we note that 
here new signals can arise from sparticle cascade decays, originating 
from a squark or a gluino. These cascades generally end in the lightest 
supersymmetric particle, which can decay to jets plus a charged lepton 
(or jets plus a neutrino, i.e. MET) when R-parity is violated. 
Accordingly, we focus on the final state with a dilepton and multiple 
jets (arising in the cascade decays) which is the analogue, in our case, 
of the jets + missing energy channel known to be the most promising of 
supersymmetry signals when R-parity is conserved. We demonstrate that 
this channel would show a large excess over the Standard Model 
background, both in the case when the gluino decays into squarks and 
when the squarks decay into gluinos. This will happen irrespective of 
whether a resonance can be seen or not.

In case a resonance is not seen, and we observe only the dilepton + jets 
signal, it may be argued, even if for the sake of argument alone, that 
some simple extensions of the Standard Model could also, perhaps, be 
tailored to produce this signal. However, supersymmetry provides another 
characteristic signal, which is enhanced if R-parity is violated, and 
this is the appearance of pairs, triads or even quartets of like-sign 
leptons in cascade decays of sparticles. Such signals have little or no 
Standard Model background. While same-sign lepton pairs appear even in 
supersymmetric models where R-parity is conserved, triads and quartets 
can only appear if R-parity is violated. We study same-sign lepton 
signals, therefore, and demonstrate that for our model these 
combinations appear in the approximate ratio 100:10:1 respectively. The 
actual number will, of course, depend most crucially on the masses of 
the squarks and the gluino, and will, therefore, depend on the point in 
the supersymmetric parameter space. We explore this parameter space and 
find out the limits to which same-sign lepton pairs, triads and quartets 
can be seen. Within these limits, one should expect the resonance or 
dilepton + jets signals to be supplemented by the observation of such 
same-sign leptons. We also show that the last-mentioned category of 
signals could make their appearance even at the 7~TeV run of the LHC, in 
the most favourable part of the parameter space.

R-parity violation is admittedly not the most popular supersymmetric 
option, probably because it causes the lightest supersymmetric particle 
to decay and hence has no explanation for dark matter. However, we 
demonstrate in this work that it can not only correlate some of the 
present and projected low-energy data comprehensively, but it can also 
predict rather spectacular signals at the LHC --- even, perhaps, in the 
early runs. If some such signals, e.g. same-sign triads, are actually 
observed, it would then be really necessary to consider the 
R-parity-violating version of supersymmetry in all seriousness, at the 
cost of leaving the dark matter problem completely open to speculation. 
This is an option which must be left to the future to decide, when the 
LHC data become available.
 
%%%%%%%%%%%%%%%%%%%%%%%%%%%%%%%%%%%%%%%%%%%%%%%%%%%%%%%%%%%%%%%%%%%%%%%%%%%%%% 
\begin{quotation}
\def\baselinestretch{1.0}
\noindent\small {\it Acknowledgments}: The authors acknowledge 
discussions with M.~Guchait and G.~Majumdar, both of the CMS 
Collaboration. BB would like to thank the Saha Institute of Nuclear 
Physics, Kolkata, for hospitality while some of the work was being done. 
GB would like to thank the Department of Theoretical Physics, Tata 
Institute of Fundamental Research, Mumbai, for an Adjunct Faculty 
position and for hospitality during a major part of this work.
\def\baselinestretch{1.2}
\end{quotation}
\normalsize

%%%%%%%%%%%%%%%%%%%%%%%%%%%%%%%%%%%%%%%%%%%%%%%%%%%%%%%%%%%%%%%%%%%%%%%%%%%%%%

%%%%%%%%%%%%%%%%%%%%%%%%%%%%%%%%%%%%%%%%%%%%%%%%%%%%%%%%%%%%%%%%%%%%%%%%%%%%%%

\begin{thebibliography}{99}

\bibitem {Altarelli} For comprehensive reviews, see G.~Altarelli, 
CERN preprint 
CERN-PH-TH/2008-085 (2008) [arXiv:0805.1992] and CERN preprint 
CERN-PH-TH/2010-249 (2010) [arxiv:1010.5637].

\bibitem{Quigg} This is nicely discussed by C.~Quigg, Fermilab preprint 
FERMILB-PUB-09/2307 (2009) [arXiv:0905.3187].

\bibitem{rpar} G.~R.~Farrar and P.~Fayet, Phys.\ Lett.\ B {\bf 76}, 575 
(1978); S.~Weinberg, Phys.\ Rev.\ D {\bf 26}, 287 (1982); N.~Sakai and 
T.~Yanagida, Nucl.\ Phys.\ B {\bf 197}, 533 (1982); C.~S.~Aulakh and 
R.~N.~Mohapatra, Phys.\ Lett.\ B {\bf 119}, 136 (1982).

\bibitem{Ross}
L.~E.~Ibanez and G.~G.~Ross, Phys.\ Lett.\ B {\bf 260}, 291 (1991); 
Nucl.\ Phys.\ B {\bf 368}, 3 (1992).

\bibitem{Bhattacharyya:2009hb} G.~Bhattacharyya, K.~B.~Chatterjee and 
S.~Nandi, Nucl.\ Phys.\ B {\bf 831}, 344 (2010).

\bibitem{RPV_LHC} D.~K.~Ghosh, R.~M.~Godbole and S.~Raychaudhuri, TIFR 
preprint TIFR-TH-99-12 (1999) [hep-ph/9904233]; S.~P.~Das, A.~Datta and 
S.~Poddar, Phys.\ Rev.\ D {\bf 73}, 075014 (2006); V.~M.~Abazov {\it et 
al.}  [D0 Collaboration], Phys.\ Lett.\ B {\bf 638}, 441 (2006); 
A.~Datta and S.~Poddar, Phys.\ Rev.\ D {\bf 75}, 075013 (2007); A.~Datta 
and S.~Poddar, Phys.\ Rev.\ D {\bf 79}, 075021 (2009); J.~M.~Butterworth 
{\it et al.}, Phys.\ Rev.\ Lett.\ {\bf 103}, 241803 (2009); N.~Desai and 
B.~Mukhopadhyaya, JHEP {\bf 1010}, 060 (2010); K.~Ghosh, S.~Mukhopadhyay 
and B.~Mukhopadhyaya, JHEP {\bf 1010}, 096 (2010); S.~L.~Chen {\it et 
al.}, [arXiv:1011.2214] (2010).

\bibitem{pdg} K. Nakamura {\it et al.} (Particle Data Group), J.\ Phys.\ 
G {\bf 37}, 075021 (2010) [http://pdg.lbl.gov].

\bibitem{Bona:2007qt} M.~Bona {\it et al.}, [arXiv:0709.0451 (hep-ex)].

\bibitem{reviews} For reviews see, G.~Bhattacharyya, Nucl.\ Phys.\ 
Proc.\ Suppl.\ {\bf 52A}, 83 (1997); G.~Bhattacharyya, 
[arXiv:hep-ph/9709395]; H.~K.~Dreiner, [arXiv:hep-ph/9707435]; 
M.~Chemtob, Prog.\ Part.\ Nucl.\ Phys.\ {\bf 54}, 71 (2005); R.~Barbier 
{\it et al.}, Phys.\ Rept.\ {\bf 420}, 1 (2005).

\bibitem{Kao:2009fg} For a recent update on single RPV couplings, 
see Y.~Kao and T.~Takeuchi, [arXiv:0910.4980 (hep-ph)].

\bibitem{Bhattacharyya:1995pq} G.~Bhattacharyya and D.~Choudhury, Mod.\ 
Phys.\ Lett.\ A {\bf 10}, 1699 (1995).

\bibitem{gb} G.~Bhattacharyya, H.~V.~Klapdor-Kleingrothaus and H.~Pas, 
Phys.\ Lett.\ B {\bf 463}, 77 (1999); S.~Rakshit, G.~Bhattacharyya and 
A.~Raychaudhuri, Phys.\ Rev.\ D {\bf 59}, 091701 (1999); A.~Abada, 
G.~Bhattacharyya and M.~Losada, Phys.\ Rev.\ D {\bf 66}, 071701 (2002). 
This is only a partial list. See \cite{reviews} for other references on 
neutrino mass generation by RPV couplings.

\bibitem{Dey:2008ht} P.~Dey {\it et al.}, JHEP {\bf 0812}, 100 (2008).

\bibitem{Nandi:2006qe} S.~Nandi and J.~P.~Saha, Phys.\ Rev.\ D {\bf 74}, 
095007 (2006).

\bibitem{Kundu:2008ui} A.~Kundu and S.~Nandi, Phys.\ Rev.\ D {\bf 78}, 
015009 (2008); A possible contribution from $\lamp$ product couplings 
has also been mentioned in A.~G.~Akeroyd and S.~Recksiegel, Phys.\ 
Lett.\ B {\bf 554}, 38 (2003).

\bibitem{Rosner:2010ak} For a review of all constraints on $f_{D_s}$, 
see J.~L.~Rosner and S.~Stone, [arXiv:1002.1655 (hep-ex)], and 
references therein.

\bibitem{:2008sq} B.~I.~Eisenstein {\it et al.}  [CLEO Collaboration], 
Phys.\ Rev.\ D {\bf 78}, 052003 (2008).

\bibitem{:2007ws} K.~Abe {\it et al.}  [Belle Collaboration], Phys.\ 
Rev.\ Lett.\ {\bf 100}, 241801 (2008).

\bibitem{Naik:2009tk} P.~Naik {\it et al.}  [CLEO Collaboration], Phys.\ 
Rev.\ D {\bf 80} 112004 (2009).

\bibitem{Dobrescu:2008er} B.~A.~Dobrescu and A.~S.~Kronfeld, Phys.\ 
Rev.\ Lett.\ {\bf 100}, 241802 (2008); R.~Benbrik and C.~H.~Chen, Phys.\ 
Lett.\ B {\bf 672}, 172 (2009); I.~Dor\v sner {\it et al.}, 
[arXiv:0906.5585 (hep-ph)].

\bibitem{gminus2} F.~Jegerlehner and A.~Nyffeler, Phys.\ Rept.\ {\bf 
477}, 1 (2009). For a recent update, see J.~Prades, [arXiv:0909.2546 
(hep-ph)].

\bibitem{cascades} H.~Baer, V.~D.~Barger, D.~Karatas and X.~Tata,
Phys. Rev. D {\bf 36}, 96 (1987) ; H.~Baer, X.~Tata and J.~Woodside, Phys.\ Rev.\ D
{\bf 41}, 906 (1990); R.~Barbieri, F.~Caravaglios, M.~Frigeni and M.~L.~Mangano,
Nucl.\ Phys.\ B {\bf 367}, 28 (1991); R.~M.~Barnett,J.~F.~Gunion and H.~E.~Haber,
Phys.\ Lett.\ B {\bf 315}, 349 (1993); H.~K.~Dreiner, M.~Guchait and D.~P.~Roy,
Phys.\ Rev.\ D {\bf 49}, 3270 (1994).

\bibitem{SUSYsearch} For a recent review, see, for example, N. \"Ozt\"urk, 
for the ATLAS and CMS Collaborations, [arXiv:0910.2964 (hep-ph)].

\bibitem{JetB} B.~Bhattacherjee {\it et al}, Phys.\ Rev.\ D {\bf 81}, 
035021 (2010).

\bibitem{BargerHan} V.~D.~Barger, G.~F.~Giudice and T.~Han, Phys.\ Rev.\ 
D {\bf 40}, 2987 (1989).

\bibitem{Tonelli} G.~Tonelli (CMS Collaboration), {\it private 
communication}.

\bibitem{leptoquark} B.~Dion {\it et al.}, Eur.\ Phys.\ J.\ C {\bf 2}, 
497 (1998); O.~J.~P.~\'Eboli, R.~Z.~Funchal and T.~L.~Lungov, 
Phys.\ Rev.\ D {\bf 57}, 1715 (1998); A.~Belyaev {\it et al}, Phys.\ Rev.\ D 
{\bf 59}, 075007 
(1999); S.~Abdullin and F.~Charles, Phys.\ Lett.\ B {\bf 464}, 223 
(1999); M.~Kramer {\it et al.}, Phys.\ Rev.\ D {\bf 71}, 057503 (2005); 
P.~Fileviez Perez {\it et al.}, Nucl.\ Phys.\ B {\bf 819}, 139 (2009); 
V.~M.~Abazov {\it et al.}  [D0 Collaboration], Phys.\ Lett.\ B {\bf 
693}, 95 (2010).

\bibitem{HOD} B.~Mukhopadhyaya and S.~Mukhopadhyay, Phys.\ Rev.\ D {\bf 
82}, 031501 (2010).

\bibitem{Pythia} T.~Sj{\o}strand, S.~Mrenna and P.~Z.~Skands, JHEP {\bf 
0605}, 026 (2006).

\bibitem{Suspect} A.~Djouadi, J.L.~Kneur and G.~Moultaka, CERN preprint 
CERN-TH/2002-32 (2002) [hep-ph/0211331].

\bibitem{ATLAS:b_tag} S.~Cucciarelli (CMS Collaboration), Nuclear 
Physics B (Proc. Suppl.) {\bf 120}, 190 (2003); M.~Lehmacher (ATLAS 
Collaboration) [arXiv:0809.4896 (hep-ex)].

\bibitem{mSUGRA} 
E.~Cremmer et al., Phys.\ Lett.\ B {\bf 79}, 231 (1978); Nucl.\ Phys.\ B
{\bf 147}, 105 (1979); R.~Barbieri, S.~Ferrara,
and C.~A.~Savoy, Phys.\ Lett.\ B {\bf 119}, 343 (1982); A.~H.~Chamseddine, 
R.~L.~Arnowitt, and P.~Nath, Phys.\ Rev.\ Lett.\ {\bf 49}, 970 (1982); 
L.~J.~Hall, J.~D.~Lykken, and S.~Weinberg, Phys.\ Rev.\ D {\bf 27}, 2359
(1983); P.~Nath, R.~L.~Arnowitt, and A.~H.~Chamseddine, Nucl.\ Phys.\ B 
{\bf 227}, 121 (1983); N.~Ohta, Prog.\ Theor.\ Phys.\ {\bf 70}, 542 (1983).
For more recent analyses, see, for example, O.~Buchm\"uller et. al, Eur.\
Phys.\ J.\ C {\bf 71}, 1583 (2011) and references theriein.

\bibitem{Allanach}
B.~C.~Allanach, A.~Dedes and H.~K.~Dreiner, Phys.\ 
Rev.\ D {\bf 69}, 115002 (2004); [Erratum-ibid.\ D {\bf 72}, 079902 
(2005)].

\bibitem{SSL}
H.~Baer, A.~Lessa and H.~Summy, Phys.\ Lett.\ B {\bf 674}, 49 (2009);
H.~Baer, V.~Barger, A.~Lessa and X.~Tata, JHEP {\bf 0909}, 063 (2009);
J.~Edsjo, E.~Lundstrom, S.~Rydbeck and J.~Sjolin, JHEP {\bf 1003}, 054 (2010). 

\end{thebibliography}
\end{document}